%% file: Formatting-Instructions-LaTeX-2026.tex
\documentclass[letterpaper]{article} 
\usepackage{aaai2026}  
\usepackage{times}  
\usepackage{helvet}  
\usepackage{courier}  
\usepackage[hyphens]{url}  
\usepackage{graphicx} 
\usepackage{natbib}  
\usepackage{caption} 
\usepackage{algorithm}
\usepackage{algorithmic}
\usepackage{amsfonts}       

\usepackage{amsmath}
\usepackage{amsthm}
\usepackage{booktabs}
\usepackage{tabularx}
\usepackage{makecell}
\usepackage{algorithm}
\usepackage{multirow}
\usepackage{booktabs}
\usepackage{pdfpages}
\usepackage{graphicx}
\usepackage{bbding}
\usepackage{multibib}

\usepackage{array}
\usepackage{listings}

\usepackage{algorithmic}
\usepackage{pifont}
\usepackage{subfig}
\usepackage{colortbl}
\usepackage{paralist}
\usepackage{hhline}

\newtheorem{theorem}{Theorem}
\newtheorem{proposition}{Proposition}
\newtheorem*{proposition*}{Proposition}

\newtheorem{definition}{Definition}

\usepackage{tikz}
\newcommand{\blacklabel}[1]{%
	\begin{tikzpicture}[]
		\node[draw=none, fill=black, text=white, inner sep=0.35pt, font=\small, circle]{#1};
	\end{tikzpicture}
}

\usepackage{newfloat}
\usepackage{listings}
\DeclareCaptionStyle{ruled}{labelfont=normalfont,labelsep=colon,strut=off} 
\floatstyle{ruled}
\newfloat{listing}{tb}{lst}{}
\floatname{listing}{Listing}
\title{Class-feature Watermark: A Resilient Black-box Watermark Against Model Extraction Attacks}
\author {
	Yaxin Xiao\textsuperscript{\rm 1}, 
	Qingqing Ye\textsuperscript{\rm 1}, 
	Zi Liang\textsuperscript{\rm 1}, 
	Haoyang Li\textsuperscript{\rm 1}, 
	RongHua Li\textsuperscript{\rm 1}, 
	Huadi Zheng\textsuperscript{\rm 2}, 
	Haibo Hu
	\thanks{Corresponding author: haibo.hu@polyu.edu.hk}\textsuperscript{\rm 1,\rm 3}
}
\affiliations {
	\textsuperscript{\rm 1}Department of Electronical and Electronic Engineering, The Hong Kong Polytechnic University\\
	\textsuperscript{\rm 2}Huawei Technologies Co., Ltd.\\
	\textsuperscript{\rm 3}Research Centre for Privacy and Security Technologies in Future Smart Systems\\
	20034165r@connect.polyu.hk, qqing.ye@polyu.edu.hk, $\{$zi1415926.liang, hao-yang9905.li, cory-ronghua.li$\}$@connect.polyu.hk, zhenghuadi@huawei.com, haibo.hu@polyu.edu.hk 
}

\usepackage{bibentry}

\begin{document}

\maketitle
\begin{abstract}
\vspace{-0.1cm}
Machine learning models constitute valuable intellectual property, yet remain vulnerable to model extraction attacks (MEA), where adversaries replicate their functionality through black-box queries. Model watermarking counters MEAs by embedding forensic markers for ownership verification. Current black-box watermarks prioritize MEA survival through representation entanglement, yet inadequately explore resilience against sequential MEAs and removal attacks. Our study reveals that this risk is underestimated because existing removal methods are weakened by entanglement. To address this gap, we propose \textbf{W}atermark \textbf{R}emoval attac\textbf{K} (WRK), which circumvents entanglement constraints by exploiting decision boundaries shaped by prevailing sample-level watermark artifacts. WRK effectively reduces watermark success rates by $\geq$88.79\% across existing watermarking benchmarks.

For robust protection, we propose \textbf{C}lass-\textbf{F}eature \textbf{W}atermarks (CFW), which improve resilience by leveraging class-level artifacts. CFW constructs a synthetic class using out-of-domain samples, eliminating vulnerable decision boundaries between original domain samples and their artifact-modified counterparts (watermark samples). CFW concurrently optimizes both MEA transferability and post-MEA stability. Experiments across multiple domains show that CFW consistently outperforms prior methods in resilience, maintaining a watermark success rate of $\geq70.15\%$ in extracted models even under the combined MEA and WRK distortion, while preserving the utility of protected models. 
\end{abstract}

\begin{links}
	\vspace{-0.305cm}
    \link{Code}{https://github.com/ClassFeatureWatermark/ClassFeatureWatermark}
    	\vspace{-0.01cm}
\end{links}

\input{introduction}
\input{definition}
\input{resilience_of_entangled}
\input{wrk}
\input{wrk_experiment}
\input{method}
\input{experiments}
\input{related_works}

\vspace{-0.258cm}
\section{Conclusion} 
\vspace{-0.10cm}
\label{sec:conclusion}
This paper first exposes resilience gaps in black-box watermarks against model extraction attacks (MEA). While existing removal methods are often constrained by the representation entanglement deliberately designed in these watermarks, our Watermark Removal attacK (WRK) overcomes this entanglement barrier by exploiting decision boundaries shaped by their sample-level artifacts. To address the vulnerability exposed by WRK, we introduce Class-Feature Watermarks (CFWs) which shift to class-level artifacts for enhanced resilience and further optimize representation entanglement and stability during MEA. CFW achieves watermark success rates of $\geq70.15\%$ under combined MEA and WRK attacks across domains, significantly outperforming prior methods in resilience.

\section{Acknowledgments}
This work was supported by the National Natural Science Foundation of China (Grant No: 92270123), and the Research Grants Council (Grant No:  15209922, 15210023, 15224124 and 15207725), Hong Kong SAR, China. 


\input{Formatting-Instructions-LaTeX-2026.bbl}
\newpage
\appendix
\input{appendix}

\end{document}

%% file: introduction.tex
\vspace{-0.42cm}
\section{Introduction}
\vspace{-0.05cm}
\label{sec:intro}
Developing machine learning (ML) models requires substantial investments in data collection, hyper-parameter tuning, and computation resources. To lower these barriers, Machine Learning as a Service (MLaaS) platforms like Google AutoML~\cite{google-automl}, Microsoft Azure ML~\cite{microsoft-azure-ml}, and Amazon SageMaker~\cite{amazon-sagemaker} offer ML inference services for applications like speech recognition~\cite{alharbi2021automatic} and medical analysis~\cite{chen2024expanding}. However, such models face the threat of model extraction attacks (MEA)~\cite{tramer2016stealing,liang25yes}, where adversaries use black-box queries to replicate model functionality~\cite{zhang25mer}. These stolen models can be monetized, which infringes on the intellectual property of model owners and highlights the need for ownership protection.

Black-box model watermarking~\cite{jia2021entangled} has emerged as a promising forensic solution to protect ownership under MEA. It embeds a watermark task into the protected model. During MEA, the stolen model inherits this task, providing ownership proof. Recent approaches~\cite{jia2021entangled,lv2024mea} employ backdoor-based watermarks that modify domain inputs with artifacts and assign new labels to create watermark samples. The model’s behavior on these samples serves as the forensic marker. These methods ensure stable transferability in MEA through entanglement with domain tasks in representation spaces. 

This focus on MEA survival creates a critical vulnerability: watermarks remain exposed to removal attacks \textit{after} model extraction. Such sequential threats are increasingly practical, as advanced removal techniques~\cite{zhu2023selective,zheng2022data,li2023reconstructive,zhang2024exploring} continue to emerge. If adversaries can strip watermarks post-extraction, ownership verification fundamentally fails, rendering defenses useless precisely when needed most.

\textbf{This paper investigates watermark resilience against sequential MEA and removal attacks. We expose a critical vulnerability:} Existing evaluation underestimates removal attack risks because existing techniques, including trigger inversion~\cite{wang2019neural,aiken2021neural}, neuron pruning~\cite{liu2018fine,zheng2022data}, and learning-induced forgetting~\cite{zhu2023selective,li2021neural}, fail against entangled watermarks. These ill-suited methods compromise resilience assessments, creating a false sense of security.

To provide a more reliable assessment, we propose \textbf{W}atermark \textbf{R}emoval attac\textbf{K} (WRK), a removal method that explicitly targets entangled black-box watermarks. WRK reshapes the decision boundary to disrupt regions likely to contain watermarks. Current watermarking paradigms construct watermark samples by injecting artifacts into domain inputs and assigning new labels, which positions the decision boundary between the original and modified samples. WRK exploits this geometric vulnerability to disrupt watermark regions. It further incorporates an objective to shift output-layer parameters~\cite{min2023towards} to prompt disentanglement. By specifically targeting the backdoor-style constructions, WRK enables more rigorous stress-testing of watermark resilience.

In the second part of this paper, to address the low resilience of backdoor-style watermarks, we introduce \textbf{C}lass-\textbf{F}eature \textbf{W}atermarks (CFW). Existing watermarks rely on sample-specific artifacts, which expose watermark regions and make them vulnerable to WRK. CFW addresses this vulnerability with a synthetic watermark class created by labeling diverse out-of-domain (OOD) samples as a unified class. In other words, it maintains essential task distinctiveness at the class level to prevent false ownership claims.

However, deploying a crafted class as a model watermark is insufficient to defend against MEA due to two key challenges. First, the task must exhibit representation entanglement (RE) with domain tasks to ensure transferability during MEA. We address this by introducing a quantitative metric to guide RE during watermark embedding, which also improves resilience by making the watermark less susceptible to learning-induced removal and pruning. Second, to remain resilient in the extracted model, CFW must preserve clustering among its samples in the representation space. Yet, feature variance among CFW samples causes uneven distortion during MEA, which disperses representations. To preserve CFW stability after MEA, we regularize pairwise distances among CFW representations to promote compact clustering.

In summary, this paper makes the following contributions:
\vspace{-0.5cm}
\begin{itemize}[$~~\bullet$]
	\item We theoretically establish a positive correlation between watermark transferability in MEA and their resilience to removal attacks, both of which are linked to the underlying property of representation entanglement.
	
	\item We identify the overlooked vulnerability of entangled watermarks, and propose \textbf{W}atermark \textbf{R}emoval attac\textbf{K} (WRK), which disentangles watermarks from domain tasks by reshaping the associated decision boundaries.
	
	\item We propose Class-Feature Watermarks (CFW), a resilient watermarking approach that leverages class-level artifacts to resist WRK. To ensure both effectiveness and resilience during MEA, we optimize its representation entanglement and stability during embedding.
	
	\item Comprehensive evaluations show that WRK reduces watermark success rates by $\geq$88.79\% across existing black-box benchmarks, whereas eight prior removal methods fail against entangled watermarks. CFW achieves superior resilience, maintaining $\geq$70.15\% watermark success rate under combined MEA and WRK attacks across domains while preserving the utility of protected models.
\end{itemize}


%% file: definition.tex
\vspace{-0.10cm}
\section{Problem Definition}
\vspace{-0.02cm}
\label{sec:problem}
\noindent\textbf{Machine Learning (ML) Notation.} 
Since model extraction attacks (MEAs) primarily target classifiers~\cite{tramer2016stealing,xiao2022mexmi}, we consider a $K$-class classification model $F_{\theta}$ (abbreviated as $F$) with parameters $\theta$. $F$ maps inputs $\mathbf{x} \in {X} \subseteq \mathbb{R}^d$ to discrete labels $y \in {Y} = \{1,\dots,K\}$. The model comprises $L$ sequential neural network layers, where layer $1$ is the input layer and layer $L$ the output layer. The representation at layer $l$ is denoted $F_{\theta^l}(\mathbf{x})$, with $\theta^l$ representing parameters from layers $1$ to $l$. The final layer outputs logits $=F(\mathbf{x}) \in \mathbb{R}^K$. These logits are normalized via softmax to produce class probabilities, with the predicted label given by $\hat{y} = \arg\max_{k \in {Y}} F(\mathbf{x})_k$.
\vspace{0.02cm}

\noindent\textbf{Model Extraction Attacks (MEAs).} 
In MLaaS, models are deployed as black-box services to protect costly model development and proprietary training data~\cite{microsoft-azure-ml,google-automl}. Model extraction attacks (MEAs) threaten the intellectual property of these models by stealing functionality through queries~\cite{zheng22protecting}. 

MEA adversaries replicate victim models $F_v$ into copy models $F_s$ using queried input-output pairs, without accessing internal structures. MEA performance is evaluated by two metrics: 1) \textbf{Test accuracy} (ACC) of $F_s$ on the domain test set $D_t$; and 2) \textbf{Fidelity} (FID)~\cite{zheng19bdpl,zhang25mer}, which measures the similarity between the victim and copy models by their label agreement rate on $D_t$. To date, learning-based model extraction is the de-facto approach, where the adversary first queries $F_v$ and then trains $F_s$ using the queried results~\cite{xiao2022mexmi,pal2020activethief}.

\noindent\textbf{Threat Model.} 
We formalize the threat model as an ownership game between a defender $\mathbf{D}$, who owns the victim model $F_v$ and aims to protect its intellectual property, and an adversary $\mathbf{A}$, who attempts to steal and misuse the model. 

Specifically, the game proceeds as follows: First, $\mathbf{D}$ embeds a black-box watermark $\mathcal{W}$ into the victim model $F_v$ to later verify ownership. Watermark samples and training data are securely stored on a trusted platform with a timestamp. Then, the adversary $\mathbf{A}$ steals the model either via model extraction attacks (black-box access) or illegal downloads (white-box access), and obtains a copy model $F_s$. To evade detection, $\mathbf{A}$ attempts to remove the watermark from $F_s$ using limited domain data $D_d$ before deploying it as a competing query service. Finally, $\mathbf{D}$ queries the suspected model $F_s$ using watermark samples to obtain evidence $\mathcal{E}$ regarding the existence of $\mathcal{W}$. The game outcomes are settled as follows: if the evidence $\mathcal{E}$ confirms the presence of $\mathcal{W}$ in $F_s$, then $\mathbf{D}$ wins; otherwise, $\mathbf{A}$ wins. However, if $\mathcal{E}$ falsely indicates the watermark in an innocent model, $\mathbf{D}$ loses. 

\noindent\textbf{Problem Formulation.} 
To succeed in the ownership game, the model watermarking must meet these criteria: \blacklabel{1}\textbf{Prop. 1}. \emph{Utility Preservation}. Watermarks preserve the functionality of the victim model $F_v$ for benign users. \blacklabel{2}\textbf{Prop. 2}. \emph{High MEA transferability}. Watermarks persist in $F_s$ across diverse MEA. \blacklabel{3}\textbf{Prop. 3}. \emph{Correctness}. The watermark accurately identifies stolen models without false ownership claims. \blacklabel{4}\textbf{Prop. 4}. \emph{Removal Resilience}. Watermarks withstand removal attempts. \blacklabel{5}\textbf{Prop. 5}. \emph{Stability}. Watermark resilience in $F_s$ consists of that in $F_v$. \blacklabel{6}\textbf{Prop. 6}. \emph{Stealthiness}. Watermarks remain undetectable to adversaries.

%% file: resilience_of_entangled.tex
\vspace{-0.15cm}
\section{Theoretical Analysis of Watermark Resilience}
\vspace{-0.02cm}
\label{sec:highly_entangled}
Existing removal methods~\cite{zhu2023selective,zhang2024exploring} insufficiently evaluate the resilience of black-box watermarks designed for ownership verification in MEA. This is because they overlook representation entanglement (RE) between watermark and domain tasks~\cite{jia2021entangled}, which is essential for watermark transferability in MEA. We present a theoretical framework showing that RE also underpins resilience against removal attacks, underscoring the need for tailored removal methods to evaluate highly entangled watermarks. Next, we first introduce a metric to quantify RE, then link it to watermark resilience using Neural Tangent Kernel (NTK) theory~\cite{doan2021atheoretical}.

\vspace{-0.18cm}
\subsection{Quantifying Representation Entanglement (RE)}
\vspace{-0.048cm}
\label{subsec:repre_similarity}

Representation entanglement (RE) refers to the similarity between representations of watermark and domain tasks. It plays a key role in enabling watermark's MEA transferability~\cite{jia2021entangled, lv2024mea}, which is evaluated as the watermark success rate in the extracted model. Although RE has been widely acknowledged, a formal definition has been lacking. We address this gap by defining RE as a measurable quantity and linking it to the success of MEA.

Although RE is defined between watermark and domain tasks, it effectively reflects entanglement between watermark and query data that is assumed following the domain distribution. We argue that the minimum cosine similarity across all layers acts as a bottleneck: if the representations of the watermark and query data are orthogonal at any layer, MEA fails. 

We illustrate this with a linear model $F_\theta$ parameterized by $\theta \in \mathbb{R}^{m \times d}$, trained on dataset $D = X \times Y$, where $X \in \mathbb{R}^{b \times d}$ and $Y \in \mathbb{R}^{b \times m}$. The model satisfies $Y^\top = \theta X^\top$. During MEA, an adversary queries the model with $X_q$ and obtains $Y_q^\top = \theta X_q^\top$. Under this setting, we establish a formal failure condition as follows, and provide its proof in Appendix~A~\cite{xiao2025class}.
\vspace{-0.012cm}
\begin{theorem}[MEA Fails with Orthogonal Representations]
	\label{the:ortho}
	Given the linear model above, if all queried outputs are orthogonal to the training outputs, \emph{i.e.}, $\mathbf{y}_q \cdot \mathbf{y}^\top = 0$ for all $\mathbf{y}_q \in Y_q$, $\mathbf{y} \in Y$, then no estimated parameter $\hat{\theta}$ inferred from $X_q \times Y_q$ can satisfy $\hat{\theta} X^\top = \theta X^\top$. In this case, MEA fails to reproduce the model outputs on the original domain.
\end{theorem}
\vspace{-0.01cm} 
Theorem~\ref{the:ortho} reveals the general failure condition for MEA, which also applies when treating $X$ as watermark data. That is, orthogonal representations between the watermark and query data cause MEA to fail, while higher cosine similarity improves fidelity. Guided by this, we define the RE metric below. Experiments in Appendix~B~\cite{xiao2025class} validate its strong correlation with the watermark's MEA transferability.
	\vspace{-0.045cm}
	
\begin{definition}{(Representation Entanglement (RE) for Black-Box Model Watermarking)}
	\label{def:re}
	Let $F_\theta$ be a neural network with layers $\{L\}$ and feature maps $\psi \in \boldsymbol{\Psi}$, where $\psi: F_{\theta^l}(X) \rightarrow \mathbb{R}^{1 \times m^l}$. Given watermark data $X_w \sim D_w$ and domain data $X \sim D$, the RE is defined as:
	\vspace{-0.15cm}
	\begin{equation}
		\label{eq:o}
		\scriptsize
		\mathcal{RE}(F_\theta; X_w, X) = \inf_{l \in \{L\}, \psi \in \boldsymbol{\Psi}} \left|\frac{ \psi(F_{\theta^l}(X_w))\cdot \psi(F_{\theta^l}(X))^\top}{\|\psi(F_{\theta^l}(X_w))\|_2\|\psi(F_{\theta^l}(X))\|_2} \right|.
	\end{equation}
\end{definition}
\vspace{-0.31cm}
\subsection{Watermark Resilience Correlates with RE}
\vspace{-0.1cm}
\label{subsec:resilience_re}
This section establishes the theoretical link between watermark resilience and representation entanglement (RE). Following prior work~\cite{zhang2024exploring}, we categorize removal techniques into three types: (1) trigger reversion~\cite{xu2023towards}, (2) learning-induced forgetting~\cite{min2023towards}, and (3) neuron pruning~\cite{zheng2022data}. Trigger-reversion attacks also rely on learning-induced unlearning to complete the removal process. A watermark is considered resilient only if it withstands all three.

Prior work~\cite{zhang2024exploring} suggests that the watermark resistance to these removal methods correlates with the NTK cross-kernel norm {$\|\phi(X_w)\phi(X)^\top\|_2$}, where $X_w$ and $X$ denote the watermark and domain datasets, and {$\phi(\cdot) = \nabla_\theta F_\theta(\cdot)$} is the NTK feature map~\cite{doan2021atheoretical}. 


Assuming $F_\theta$ is a multi-layer perceptron (MLP), we show in Theorem~\ref{theo:relation_re_ntk} that this NTK norm is lower bounded by the RE defined in Equation~\ref{eq:o}.
\vspace{-0.08cm}
\begin{theorem}[Lower Bound of NTK Cross-Kernel Norm via RE]
	\label{theo:relation_re_ntk}
	Let $X_w$ and $X$ be datasets from the watermark and domain tasks, with sizes $N_w$ and $N$, respectively. Let $F_\theta$ be a neural network, specifically a multi-layer perceptron (MLP). Define $\Gamma = N_w N$. Then the following inequality holds:
			\vspace{-0.05cm}
	\begin{equation}
		\label{eq:theo_relationship}
		\scriptsize
		\|\phi(X_w)\phi(X)^\top\|_2 \ge \Gamma \cdot \mathcal{RE}(F_\theta; X_w, X).
		\vspace{-0.11cm}
	\end{equation}
\end{theorem}
Therefore, RE provides a theoretical foundation for resistance against existing removal methods. The formal proof of Algorithm~\ref{theo:relation_re_ntk} is given in Appendix~A~\cite{xiao2025class}.

%% file: wrk.tex
\vspace{-0.245cm}
\section{Watermark Removal Attack (WRK)}
\vspace{-0.04cm}
\label{sec:wrk}
Given the limitations of existing removal methods against watermarks designed for ownership verification of MEA-stolen models, we introduce \emph{Watermark Removal attacK} (WRK), which explicitly targets highly entangled black-box watermarks~\cite{jia2021entangled,lv2024mea} to evaluate their true resilience.

\noindent\textbf{High-level Solution}. 
The ideal way to remove entangled watermarks is to unlearn the watermark data from the model~\cite{graves2021amnesiac}. However, this is typically infeasible without access to watermark data, while inversion-based methods~\cite{wang2019neural,xu2023towards} largely fail due to the diversity of watermark artifacts. To overcome these limitations, WRK avoids direct inversion and instead perturbs suspicious input regions where watermark tasks are likely to reside. Existing watermarking methods~\cite{jia2021entangled,lv2024mea} embed artificial patterns into source samples and assign them to target labels. This implies that the decision boundaries lie between the source samples and the corresponding watermark samples. WRK aims to reshape these boundaries to disrupt watermark-related regions.

To implement this idea, WRK follows two key steps. The first, \textit{Boundary Reshaping}, uses large-magnitude adversarial perturbations to generate a set $X_{p}$, which targets the broad and variable nature of watermark artifacts. It assigns random labels to $X_{p}$ to form $D_p$ to enforce stronger boundary shifts between source and perturbed samples. In contrast, adversarial training (AT)~\cite{kurakin2016adversarial} applies relatively small perturbations and preserves clean labels, resulting in minimal decision movement. The second step, \textit{Feature Shifting}, resets the final layer and applies a parameter-shift regularization~\cite{min2023towards} to further decouple watermark-related features from the primary task.


\noindent\textbf{Algorithm}. The adversary $\mathbf{A}$ has (1) white-box access to the stolen copy model $F_s$, and (2) a small domain dataset $D_d\subset X\times Y$ of size $N_d$. WRK first constructs a perturbing dataset $D_p$ by sampling a subset $D_d^\prime$ of size $\rho N_d$ from $D_d$. For each $\mathbf{x}\in D_d^\prime$, adversarial noise $\delta_x$ with magnitude $\epsilon$ is computed using FGSM~\cite{kurakin2016adversarial}. Then, the adversarial sample is formed as $\tilde{x}\gets \text{Clip}(x+\delta_x)$, where the $\text{Clip}$ function ensures pixel values are valid. A random label $y \in \{1, \dots, K\}$ is assigned to form the adversarial pair. The resulting set of pairs $(\tilde{x}, y)$ forms $D_p$. Finally, WRK fine-tunes $F_s$ on $D_\text{train}=D_d \cup D_p$ to obtain $F_{s^\text{wrk}}$, with a regularization term with coefficient $\alpha$ to encourage the output layer $\theta^L$ of $F_s$ shifts from the initial weights $\theta^L_\text{ini}$. Formally, the fine-tuning loss $\mathcal{L}_\text{wrk}$ is:
\vspace{-0.155cm}
\begin{equation}
	\label{eq:wrk_loss}
	\scriptsize
	\mathcal{L}_\text{wrk}=\Big[\mathbb{E}_{(x,y)\sim D_{\text{train}}} \mathcal{L}(F_s(x), y)
	+ \alpha \langle \theta^L, \theta^L_{\text{ini}} \rangle\Big]
	\vspace{-0.075cm}
\end{equation}
The pseudo-code of WRK is provided in Appendix C~\cite{xiao2025class}.

%% file: wrk_experiment.tex
\vspace{-0.15cm}
\section{Experiments for WRK}
\vspace{-0.065cm}
\label{sec:ex_wrk}
We evaluate the resilience of black-box watermarks under WRK, and compare WRK with eight existing removal methods. Ablation studies of WRK and additional results on the Imagenette are provided in Appendix~E~\cite{xiao2025class}.

\vspace{-0.245cm}
\subsection{Overview of Experimental Setup} 
\vspace{-0.095cm}
\label{subsec:ex_wrk_setup}
We evaluate WRK on four black-box watermarking methods: EWE~\cite{jia2021entangled}, MBW~\cite{kim2023margin}, MEA-Defender (MEA-D)~\cite{lv2024mea}, and Blend~\cite{chen2017targeted}. As most methods are designed for image domains, we use CIFAR-10~\cite{krizhevsky2009learning} with ResNet18 and ImageNette~\cite{imagenette} with ResNet50 (see Appendix~E.1~\cite{xiao2025class} for ImageNette results). Two model extraction attacks, MExMI~\cite{xiao2022mexmi} and ActiveThief~\cite{pal2020activethief}, are used with query budgets of 25,000 (CIFAR-10) and 5,000 (ImageNette), respectively. Each removal method can access $5\%$ of the domain samples for CIFAR-10 and $10\%$ for ImageNette. Evaluation metrics include test accuracy (ACC), fidelity (FID), and watermark success rate (WSR). Full details on removal baselines, configurations, MEA settings, WRK hyperparameters, and metric definitions are detailed  in Appendix~D~\cite{xiao2025class}.
\vspace{-0.245cm}
\subsection{Resilience Evaluation of Existing Watermarks} 
\vspace{-0.095cm}
\label{subsec:ex_sota_compare}
We comprehensively evaluate black-box watermarks from two perspectives: their watermarking effectiveness and their resilience to removal attacks. Table~\ref{Table:compare_sta} reports the watermark success rate (WSR) on victim models and its transferability to extracted models via MEA. Table~\ref{Table:removal_attack} shows watermark removal results using WRK and existing removal baselines, with bold indicating the lowest WSR. To assess generality, WRK is also applied to two non-watermark backdoors, WaNet~\cite{tuan2021wanet} and composite~\cite{lin2020composite} backdoors, which are excluded from MEA-post watermarking analysis due to their poor transferability.

\begin{table*}
	\scriptsize
	\renewcommand{\arraystretch}{0.975}
	\centering
	\begin{tabular}{>{\centering}p{1.15cm}|p{1.15cm}>{\raggedleft\arraybackslash}p{1.25cm}|>{\raggedleft\arraybackslash}p{1.15cm}>{\raggedleft\arraybackslash}p{1.25cm}|>{\centering}p{1.25cm}>{\raggedleft\arraybackslash}p{1.15cm}>{\raggedleft\arraybackslash}p{1.2cm}|>{\raggedleft\arraybackslash}p{1.2cm}>{\raggedleft\arraybackslash}p{1.15cm}>{\raggedleft\arraybackslash}p{1.2cm}}
		\hline
		\textbf{Method}&\multicolumn{2}{c|}{\textbf{Non-Watermark Model}}&\multicolumn{2}{>{\centering}p{2.5cm}|}{\textbf{Victim Model}}&\multicolumn{3}{c|}{\textbf{Copy Model} (MEA: MExMI)}&\multicolumn{3}{c}{\textbf{Copy Model} (MEA: ActiveThief)}\\ \hline
		\textbf{Metric}/\%&ACC&WSR&ACC&WSR&{ACC}&{FID}&WSR&{ACC}&{FID}&WSR\\ \hline
		EWE&{93.55$\pm$0.19}&{17.53$\pm$2.45}&{91.98$\pm$0.18}&{99.88$\pm$0.12}&89.15$\pm$0.48&91.52$\pm$0.45&99.65$\pm$0.35&83.77$\pm$0.52&87.16$\pm$0.62&99.32$\pm$0.68\\ \hline
	    {MBW}&{93.55$\pm$0.19}&{10.00$\pm$0.00}&{73.77$\pm$0.31}&100.00$\pm$0.0&71.32$\pm$1.30&86.99$\pm$1.25&10.00$\pm$10.0&70.58$\pm$0.35&84.99$\pm$0.30&10.00$\pm$0.00\\ \hline
		{\makecell[c]{MEA-D}}&{93.55$\pm$0.19}&0.96$\pm$0.21&{85.93$\pm$0.10}&{96.50$\pm$0.20}&82.15$\pm$0.31&84.31$\pm$0.28&99.20$\pm$0.15&78.08$\pm$0.25&81.64$\pm$0.22&99.14$\pm$0.18 \\ \hline
		{Blend}&{93.55$\pm$0.19}&1.53$\pm$0.20&{93.55$\pm$0.08}&{99.85$\pm$0.15}&89.97$\pm$0.43&91.57$\pm$0.12&42.96$\pm$0.65&86.88$\pm$0.51&89.81$\pm$0.48&39.44$\pm$12.6 \\ \hline
	\end{tabular}
		\vspace{-0.18cm}
		\caption{Performance of Existing Black-box Watermarks}
	\label{Table:compare_sta}
	\vspace{-0.155cm}
\end{table*}

\begin{table*}
	\scriptsize
	\renewcommand{\arraystretch}{0.97}
	\centering
	\begin{tabular}{p{1.002cm}|>{\raggedleft\arraybackslash}p{0.92cm}>{\raggedleft\arraybackslash}p{1.03cm}|>{\raggedleft\arraybackslash}p{0.92cm}>{\raggedleft\arraybackslash}p{0.92cm}|>{\raggedleft\arraybackslash}p{0.9cm}>{\raggedleft\arraybackslash}p{0.92cm}|>{\raggedleft\arraybackslash}p{0.9cm}>{\raggedleft\arraybackslash}p{0.95cm}|>{\raggedleft\arraybackslash}p{0.9cm}>{\raggedleft\arraybackslash}p{1.02cm}|>{\raggedleft\arraybackslash}p{0.9cm}>{\raggedleft\arraybackslash}p{1.02cm}}
				\hline
			\textbf{Removal} & \multicolumn{2}{c|}{\textbf{EWE}} & \multicolumn{2}{c|}{\textbf{MBW}} & \multicolumn{2}{c|}{\textbf{MEA-D}} & \multicolumn{2}{c|}{\textbf{Blend}} & \multicolumn{2}{c|}{\textbf{WaNet}} & \multicolumn{2}{c}{\textbf{Component}} \\ \hline
		\textbf{Metrics} & ACC & WSR & ACC & WSR & ACC & WSR & ACC & WSR & ACC & WSR & ACC & WSR \\ \hline
	\end{tabular}
	\begin{tabular}{p{1.002cm}|>{\raggedleft\arraybackslash}p{0.92cm}>{\raggedleft\arraybackslash}p{1.03cm}>{\raggedleft\arraybackslash}p{0.92cm}>{\raggedleft\arraybackslash}p{0.92cm}>{\raggedleft\arraybackslash}p{0.92cm}>{\raggedleft\arraybackslash}p{0.92cm}>{\raggedleft\arraybackslash}p{0.92cm}>{\raggedleft\arraybackslash}p{0.95cm}>{\raggedleft\arraybackslash}p{0.92cm}>{\raggedleft\arraybackslash}p{0.98cm}>{\raggedleft\arraybackslash}p{0.92cm}>{\raggedleft\arraybackslash}p{1.00cm}}
		\textbf{None} 
		& 91.98$\pm$0.17; & 99.88$\pm$0.12 
		& 73.77$\pm$0.31; & 100.00$\pm$0 
		& 85.93$\pm$0.25; & 96.50$\pm$0.20 
		& 93.55$\pm$0.18; & 99.95$\pm$0.05 
		& 92.19$\pm$0.22; & 97.69$\pm$0.13 
		& 92.51$\pm$0.23; & 94.79$\pm$0.35 \\
		\textbf{NC} 
		& N/A & N/A 
		& 71.31$\pm$0.20; & \textbf{0.00}$\pm$0.00 
		& 86.23$\pm$0.15; & 47.68$\pm$12.7 
		& 92.84$\pm$0.10; & \textbf{2.22}$\pm$0.45 
		& N/A; & N/A 
		& N/A; & N/A \\
		\textbf{I-BAU} 
		& 90.35$\pm$0.18;& 99.97$\pm$0.03 
		& 73.15$\pm$0.22;& \textbf{0.00}$\pm$0.00 
		& 85.08$\pm$0.18;& 76.00$\pm$12.2 
		& 92.15$\pm$0.12; & 13.24$\pm$2.60 
		& 91.05$\pm$0.15 & 82.43$\pm$5.23 
		& 90.61$\pm$0.20 & 90.72$\pm$4.20 \\
		\textbf{BTI-DBF} 
		& 90.21$\pm$0.15; & 12.65$\pm$2.55 
		& 78.82$\pm$0.18; & 8.00$\pm$8.00 
		& 84.45$\pm$0.12; & 32.24$\pm$10.2 
		& 91.69$\pm$0.15; & 77.96$\pm$4.65 
		& 91.97$\pm$0.10; & 43.19$\pm$13.60 
		& 93.85$\pm$0.48; & 90.92$\pm$0.30 \\
		\textbf{CLP} 
		& 90.17$\pm$0.22; & 95.60$\pm$0.20 
		& 76.56$\pm$0.30; & 60.00$\pm$10.0 
		& 84.75$\pm$0.20; & 95.60$\pm$0.51 
		& 91.62$\pm$0.18; & 64.24$\pm$9.70 
		& 90.92$\pm$0.20; & 51.24$\pm$8.65 
		& 91.18$\pm$0.15; & 94.25$\pm$0.75 \\
		\textbf{FP} 
		& 89.92$\pm$0.17; & 99.99$\pm$0.01 
		& 72.50$\pm$0.25; & 80.00$\pm$10.0 
		& 83.70$\pm$0.22; & 89.54$\pm$3.23 
		& 91.52$\pm$0.20; & 89.14$\pm$0.55 
		& 91.58$\pm$0.18; & 23.12$\pm$7.70 
		& 91.95$\pm$0.12; & 94.85$\pm$0.20 \\
		\textbf{NAD} 
		& 91.80$\pm$0.10; & 99.98$\pm$0.02 
		& 71.90$\pm$0.15; & 10.00$\pm$10.0 
		& 84.75$\pm$0.15; & 74.62$\pm$11.8 
		& 91.56$\pm$0.15; & 76.43$\pm$5.50 
		& 91.12$\pm$0.15; & 70.07$\pm$5.45 
		& 90.56$\pm$0.18; & 94.06$\pm$0.35 \\
		\textbf{SEAM} 
		& 90.49$\pm$0.15; & 22.89$\pm$13.35 
		& 81.52$\pm$0.10; & \textbf{0.00}$\pm$0.00 
		& 84.05$\pm$0.10;& 64.62$\pm$7.36 
		& 91.63$\pm$0.10; & 87.64$\pm$1.45 
		& 91.80$\pm$0.12; & 33.24$\pm$6.50 
		& 90.55$\pm$0.15; & 91.45$\pm$1.25 \\
		\textbf{FST} 
		& 89.96$\pm$0.20; & 23.55$\pm$4.30 
		& 83.17$\pm$0.12; & 6.00$\pm$6.00 
		& 86.08$\pm$0.08; & \textbf{6.78}$\pm$0.30 
		& 92.48$\pm$0.08; & 57.51$\pm$6.75 
		& 91.88$\pm$0.08; & 81.01$\pm$5.40 
		& 91.07$\pm$0.20; & \textbf{3.80}$\pm$0.50 \\
		\textbf{WRK} 
		& 91.44$\pm$0.08; & \textbf{8.75}$\pm$1.75 
		& 82.50$\pm$0.15; & 4.00$\pm$6.00 
		& 85.46$\pm$0.12; & 7.64$\pm$0.28 
		& 92.40$\pm$0.07; & 9.80$\pm$2.35 
		& 92.06$\pm$0.05; & \textbf{0.56}$\pm$0.25 
		& 91.82$\pm$0.15; & 6.00$\pm$0.30 \\ \hline
	\end{tabular}
	\vspace{-0.235cm}
	\caption{Performance of WRK and benchmark removal attacks on \textbf{victim} models. Results on copy models are provided in Appendix~E~\cite{xiao2025class}.}
	\label{Table:removal_attack}
	\vspace{-0.42cm}
\end{table*}

\begin{figure}
	\centering{
		\hspace{-0.3cm}\subfloat[{\footnotesize EWE}]{
			\label{fig:curve_ewe}
			\includegraphics[width=0.382\columnwidth]{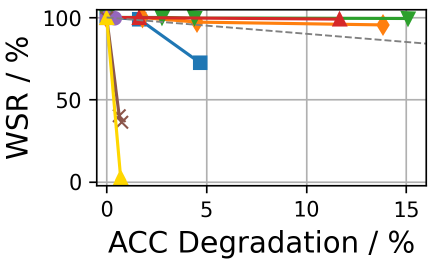}
		}\hspace{-0.2cm}
		\subfloat[{\footnotesize MEA-Defender}]{
			\label{fig:curve_mea}
			\includegraphics[width=0.382\columnwidth]{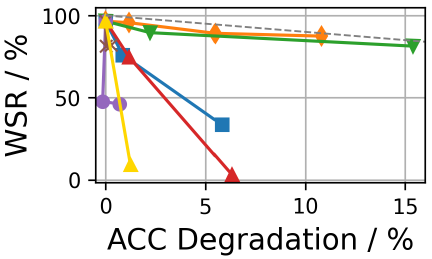}
		}\hspace{-0.15cm}
		\includegraphics[width=0.22\columnwidth]{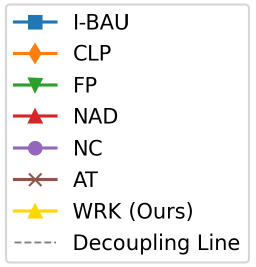}
		\vspace{-0.305cm}
		\caption{Watermark decoupling curves of victim models. On the decoupling line, ACC and WSR degrade equally.}
		\label{fig:wrk_decoupling}
	}
	\vspace{-0.38cm}
\end{figure}

\noindent\textbf{Watermarking Performance}. 
Table~\ref{Table:compare_sta} shows that EWE and MEA-Defender achieve strong MEA transferability, while Blend and MBW perform less effectively. Among them, only Blend preserves model accuracy well. In contrast, EWE causes a $1.5\%$ accuracy drop, and both MEA-Defender and MBW incur over $10\%$ degradation, mainly due to MBW’s incompatibility with data augmentation and the complexity of MEA-Defender’s loss design.

\noindent\textbf{Resilience to Removal}.
Table~\ref{Table:removal_attack} reports that no existing removal method successfully removes all watermarks and backdoors. EWE, MEA-Defender, and Blend exhibit stronger resilience than MBW, which correlates with their higher representation entanglement (recorded in Appendix~B.1). In contrast, WRK consistently reduces all WSRs below non-watermarked baselines while maintaining accuracy within $1.15\%$ of the original model. It also neutralizes all tested backdoors without trigger reversal, which highlights its broad applicability. In the case of MBW, WRK improves accuracy from $73.77\%$ to $82.50\%$ by reintroducing data augmentation during finetuning.

\noindent\textbf{Watermark Decoupling Curves}. Figure~\ref{fig:wrk_decoupling} illustrates the WSR-accuracy trade‑off curves for CIFAR‑10. Notably, WRK exhibits the steepest curve among the compared methods, achieving a substantial WSR reduction by $91.25\%$ for EWE with only a $0.54\%$ accuracy drop. While FST~\cite{min2023towards} performs comparably to WRK on MEA-Defender, it is less effective on EWE. Besides, AT~\cite{kurakin2016adversarial} and NC~\cite{wang2019neural} fall quickly at first yet stall above $35\%$ WSR, as their reliance on precise trigger/noise inversion limits removal progress.

%% file: method.tex
\vspace{-0.105cm}
\section{Class-Feature Watermark (CFW)}
\vspace{-0.06cm}
\label{sec:watermark}
Given the vulnerability of existing watermarks to WRK, we focus on designing a resilient black-box watermark that protects against MEA infringement and resists removal attacks.
\vspace{-0.625cm}
\subsection{Class-Level Artifacts for Enhanced Resilience}
\label{subsec:carrier_wm}
\vspace{-0.115cm}
The weakness of current watermarking schemes under WRK stems from their reliance on sample-level artifacts, where source samples are relabeled to unrelated target classes. This enforced label discrepancy exposes watermarks to removal through decision boundary perturbation.  

Therefore, to build a resilient watermark, we avoid sample-level artifacts and instead construct class-level artifacts. This leads to the design of the \textbf{C}lass-\textbf{F}eature \textbf{W}atermark (CFW), which maintains task distinctiveness to prevent false ownership claims. Since CFW contains no per-sample patterns, it inherently evades reversion-based methods~\cite{wang2019neural}. When optimized to enhance its representation entanglement (RE), it further withstands learning-induced forgetting~\cite{li2021neural} and neuron pruning~\cite{zheng2022data}.

CFW is formed by assigning samples from multiple out-of-domain (OOD) classes to a single watermark class. It is efficient and task-agnostic, and avoids the cost of high-resolution generation~\cite{zhu2024reliable}. Its class-level artificial distinction arises from the fact that CFW is a non-existent class. To further prevent false claims, CFW ensures that non-watermarked models neither classify these samples as a single category nor cluster their representations. Instead, they should be scattered in the output space. While this scattering occurs naturally, we can further guarantee it by selecting watermark samples using a pre-trained model akin to the step in~\cite{jia2021entangled}. Apart from this diversity constraint in the representation space, CFW introduces an additional safeguard to prevent false positive claims. When the OOD dataset is too distant from the clean task, CFW may yield high false positive rates. To address this, we measure the distance between CFW's OOD set and the domain set using RBF-MMD, normalized to $[0,1]$ via $\exp(-\text{distance}^2/\sigma^2)$. CFW claims ownership only when RBF-MMD $\geq 0.98$.

\begin{figure*}
	\centering{
		\includegraphics[width=2.0\columnwidth]{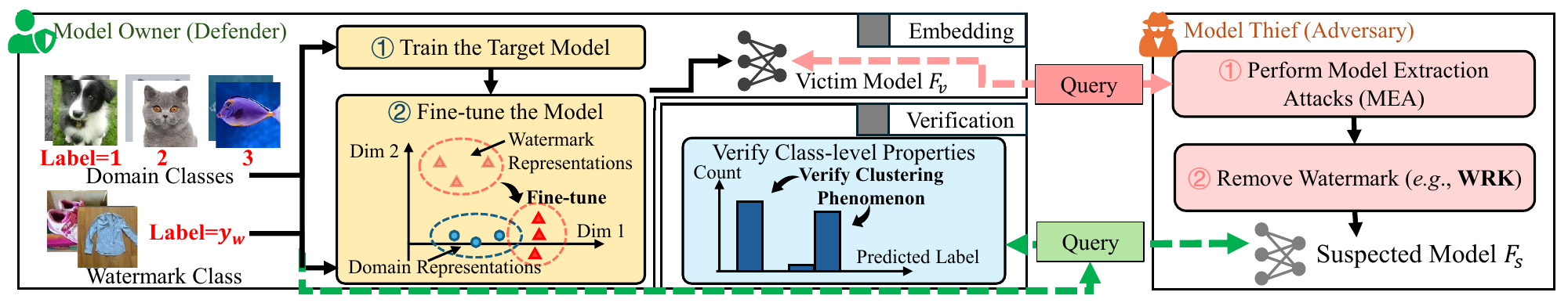}
	}
	\vspace{-0.18cm}
	\caption{Overall framework of Class-Feature Watermark (CFW).}
	\label{fig:cfw_framework}
		\vspace{-0.05cm}
\end{figure*}

\noindent\textbf{Overview of CFW}. For CFW, its \textbf{MEA transferability} guaranteed by representation entanglement (RE) and its \textbf{stability} (defined in \emph{Problem Formulation}) in MEA must be achieved. Figure~\ref{fig:cfw_framework} outlines the CFW framework which comprises two primary phases: embedding and verification. In the embedding phase, the model is first trained jointly on the domain and watermark datasets to embed the watermark (Step \ding{172}), followed by fine-tuning to enhance RE and stability (Step \ding{173}). In the verification phase, CFW uses class-level clustering behavior on the watermark task, which offers more substantial evidence of watermark presence than averaging individual sample predictions.

\vspace{-0.2cm}
\subsection{Optimize RE and Stability of CFW}
\label{subsec:cfw_ft}
\vspace{-0.135cm}
To enhance representation entanglement (RE) and stability, the fine-tuning phase employs two optimization strategies, both of which are implemented by additional loss terms. Soft Nearest Neighbor Loss (SNNL) is commonly used to increase entanglement by encouraging watermark representations to align with domain features during trigger generation~\cite{jia2021entangled}. However, it is not suitable for our CFW for two reasons. First, CFWs do not include sample-level trigger generation. Second, SNNL jointly optimizes the model and temperature parameters, making it challenging to strike a balance between RE and model utility. To overcome these issues, we design a \textbf{Rep}resentation \textbf{S}imilarity (RepS) loss that directly maximizes the cosine similarity between the average representations of watermark and domain data across layers, as guided by the metric $\mathcal{RE}$. For watermark data  $X_w\sim D_w$ and domain data $X \sim D$, the RepS loss is defined as:
\vspace{-0.156cm}
\begin{equation}
	\label{eq:loss_rep}
	\scriptsize
	\mathcal{L}_\text{RepS} =\sum\nolimits_{l=l_0}^L \left|\frac{\psi(F_{\theta^l}(X_w))\cdot\psi(F_{\theta^l}(X))^\top}{\|\psi(F_{\theta^l}(X_w))\|_2\|\psi(F_{\theta^l}(X))\|_2}\right|,
	\vspace{-0.1cm}
\end{equation}
where $l_0$ is a selected intermediate layer and $\psi:F_{\theta^l}(X)\rightarrow \mathbf{R}^{1\times m^l}$ computes the mean vectors of representations.

In addition to RE, another crucial factor for watermark robustness is the stability of watermark representations under MEA. To this end, we introduce the \textbf{Change of Distance under Distortion} loss, denoted as $L_{\text{CD}^2}$, which aims to preserve intra-class clustering of watermark samples during MEA. Starting from the approximated copy model $\hat{F}$ extracted by MEA, we define the original objective as:
\vspace{-0.108cm}
\begin{equation}
	\scriptsize
	\label{eq:loss_cd2_initial_main}
	\mathcal{L}_{\text{CD}^2} = \frac{1}{N_w^2}\sum\nolimits_{\mathbf{x}_i,\mathbf{x}_j\in X_w}|\hat{F}(\mathbf{x}_i)-\hat{F}(\mathbf{x}_j)|.
	\vspace{-0.115cm}
\end{equation}

To make this formulation tractable, we approximate $\hat{F}$ using neural tangent kernel (NTK) theory~\cite{bennani2020generalisation}. As such, the close-form $\text{CD}^2$ loss is theoretically derived from Equation~\ref{eq:loss_cd2_initial_main} as follows:
\vspace{-0.128cm}
\begin{equation}
	\scriptsize
	\label{eq:loss_cd2_final_main}
	\mathcal{L}_{\text{CD}^2} = \frac{1}{N_w^2}\sum\nolimits_{\mathbf{x}_i,\mathbf{x}_j \in D_w} \left|\left(F_{\theta^{L-1}}(\mathbf{x}_j) - F_{\theta^{L-1}}(\mathbf{x}_i)\right)V_{\theta^L}\right|,
	\vspace{-0.125cm}
\end{equation}
where $V_{\theta^L}$ denotes the right singular vectors of the domain representation matrix, \emph{i.e.}, {\small$V_{\theta^L} = \mathrm{SVD}\left(F_{\theta^{L-1}}(X)\right)$}. Complete derivation is in Appendix~A~\cite{xiao2025class}. Intuitively, $V_{\theta^L}$ captures the dominant distortion directions introduced by MEA in the representation space. Appendix~B~\cite{xiao2025class} empirically supports this by showing that MEA-induced updates align with the principal components found in $V_{\theta^L}$. Since the theory is derived for binary classification, we compute $V_{\theta^L}$ separately for each domain class in practice. Combining the above, our final fine-tuning objective becomes:
\vspace{-0.1cm}
\begin{equation}
	\label{eq:loss_obj_final_main}
	\scriptsize
	\mathcal{L} = \mathcal{L}_\text{Cri} - \lambda_1\mathcal{L}_\text{RepS} + \lambda_2\mathcal{L}_{\text{CD}^2},
	\vspace{-0.15cm}
\end{equation}
where $\mathcal{L}_\text{Cri}$ is the criterion loss that preserves both domain utility and watermark accuracy:
\vspace{-0.1cm}
\begin{equation}
	\scriptsize
	\mathcal{L}_\text{Cri} = \frac{1}{N}\sum_{\mathbf{x}\in X} \mathcal{L}(F(\mathbf{x}), F_\text{1}(\mathbf{x})) + \frac{1}{N_w}\sum_{(\mathbf{x}, y)\in D_w} \mathcal{L}(F(\mathbf{x}), y),
	\vspace{-0.16cm}
\end{equation}
and $\lambda_1, \lambda_2 > 0$ are weighting factors for the auxiliary terms. The computation of $\mathcal{L}$ incurs complexity of $O(B_w^2DK)$, where $B_w$ is the batch size of watermark samples and $D$ is the dimension of the $L$-layer output. This complexity is minimal and incurs negligible overhead, as $B_w$ is typically much smaller than the domain batch size.

\noindent\textbf{Resilience Bonus of $\text{CD}^2$ Optimization}. Beyond improving post-MEA resilience, ${\text{CD}^2}$ optimization also enhances robustness against learning-induced removal, because the removal-post model $\hat{F}$ shares the same NTK-form approximation with the MEA-post model. As a result, minimizing $\mathcal{L}_{\text{CD}^2}$ preserves intra-class cohesion not only during extraction but also in the learning-induced removal process.

\vspace{-0.195cm}
\subsection{Verify CFW with Intra-class Clustering}
\vspace{-0.095cm}
\label{sec:verification}
Watermark verification is a hypothesis test with two possible outcomes: Hypothesis 0 (H0) assumes the model is watermarked, while Hypothesis 1 (H1) asserts it is not. Existing methods~\cite{jia2021entangled,lv2024mea} typically apply t-tests and use the watermark success rate (WSR) as the test statistic~\cite{jia2021entangled}. However, these approaches ignore the class-level structure in CFWs. Given their enhanced stability, our verification method prioritizes group clustering~\cite{jin2023explaining} over individual predictions. Next, we explore why clustering provides a more resilient verification evidence than prediction accuracy (\emph{i.e.}, watermark success rate, WSR).

\noindent\textbf{Clustering is a Resilient Evidence for Watermark Presence.}
\label{subsubsec:ex_clustering}
To compare the resilience of intra-class clustering and WSR, we apply the WRK attack to class-feature watermarks. This attack perturbs the sample-wise artifacts and leverages learning-induced forgetting, making it a strong testbed for resilience.

\begin{figure}
	\centering{
		\subfloat[{\scriptsize CFW}]{
			\includegraphics[width=0.275\columnwidth]{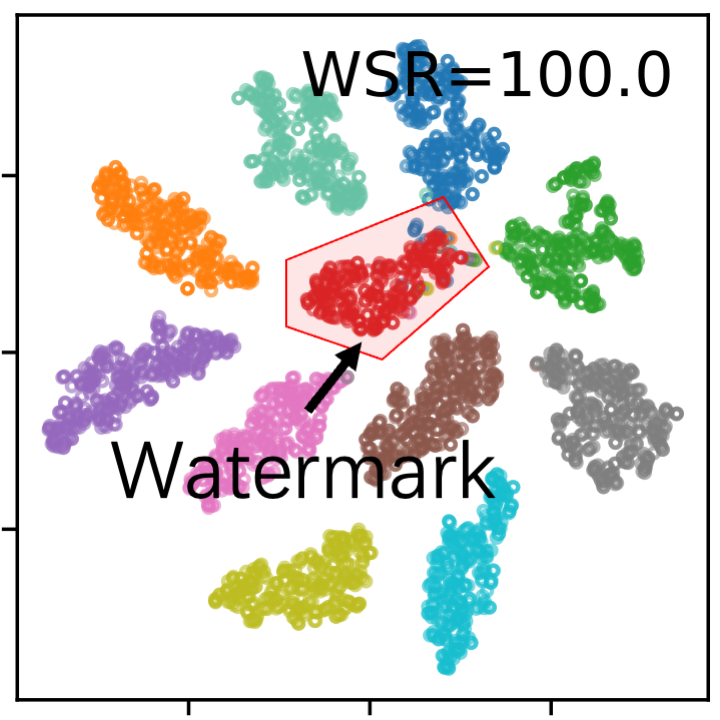}
		}\quad
		\subfloat[{\scriptsize WRK-attacked CFW}]{
			\includegraphics[width=0.275\columnwidth]{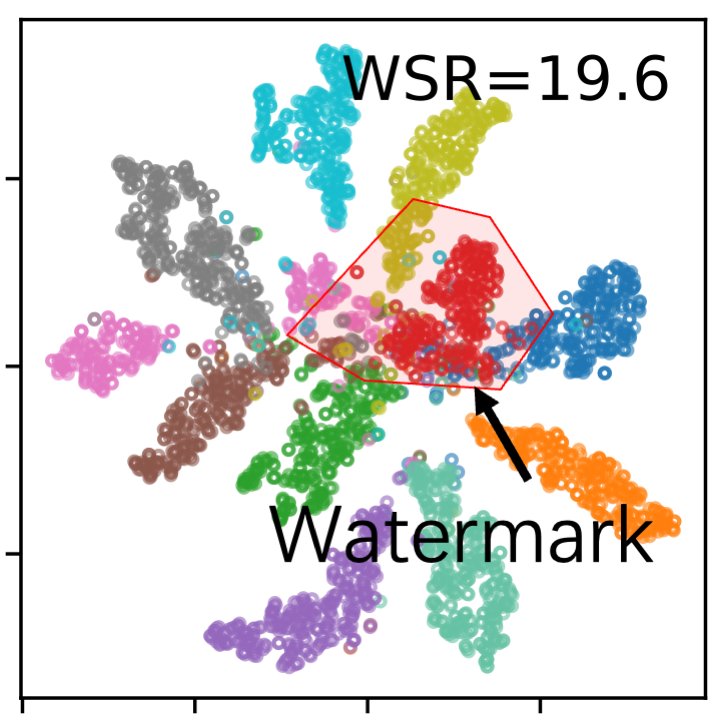}
			\label{subfig:cfw_ft}
		}\quad
		\subfloat[{\scriptsize No watermark}]{
			\includegraphics[width=0.275\columnwidth]{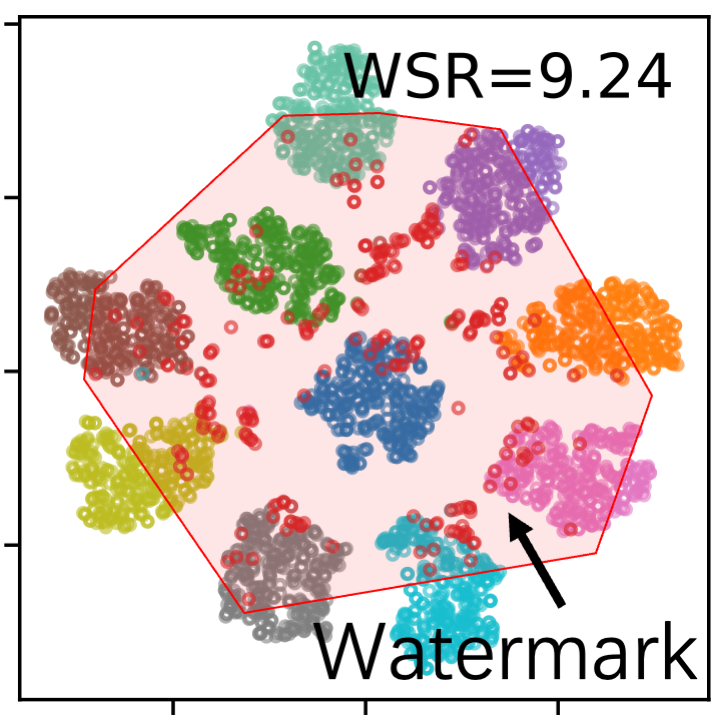}}
	}
    \vspace{-0.33cm}
	\caption{Visualized representations of the last hidden layer.}
	\vspace{-0.26cm}
	\label{fig:cf_rep}
\end{figure}

\noindent\textbf{Experiments}. Experiments are conducted on CIFAR-10 with ResNet-18, using 250 out-of-domain (OOD) samples from CIFAR-20 to create the watermark task, whose clustering is optimized with $\text{CD}^2$. In this setting, CFW shows a weak correlation between WSR and clustering, making it ideal for our comparison. Figure~\ref{fig:cf_rep}  uses t-SNE~\cite{van2008visualizing} to visualize the final-layer representations and WSR (see Equation~\ref{eq:wsr}) for three models: the original watermarked model, the WRK-attacked watermarked model, and a non-watermarked model. Figure~\ref{subfig:cfw_ft} shows that despite WSR dropping from $100\%$ to $19.60\%$, representations remain well-clustered, indicating clustering is more resilient than WSR.

\noindent\textbf{Verify with Clustering in Label-only Access.} 
Since clustering is more resilient than WSR, we propose to verify CFW by checking for clustering. However, in practice, the verifier may only have \textbf{label-only} access. In this case, we observe whether label distributions imply clustering in the watermark class. Thanks to the stability enforced by $\text{CD}^2$ optimization, we infer that its logits are still highly consistent during removal. As a result, label distributions may reveal clustering patterns even when WSR suggests failure. This phenomenon is referred to as label clustering.

\begin{figure}
	\centering{
		\subfloat[{\footnotesize WRK-Attacked CFW}]{
			\label{fig:cf_label_cluster}
			\includegraphics[width=0.401\columnwidth]{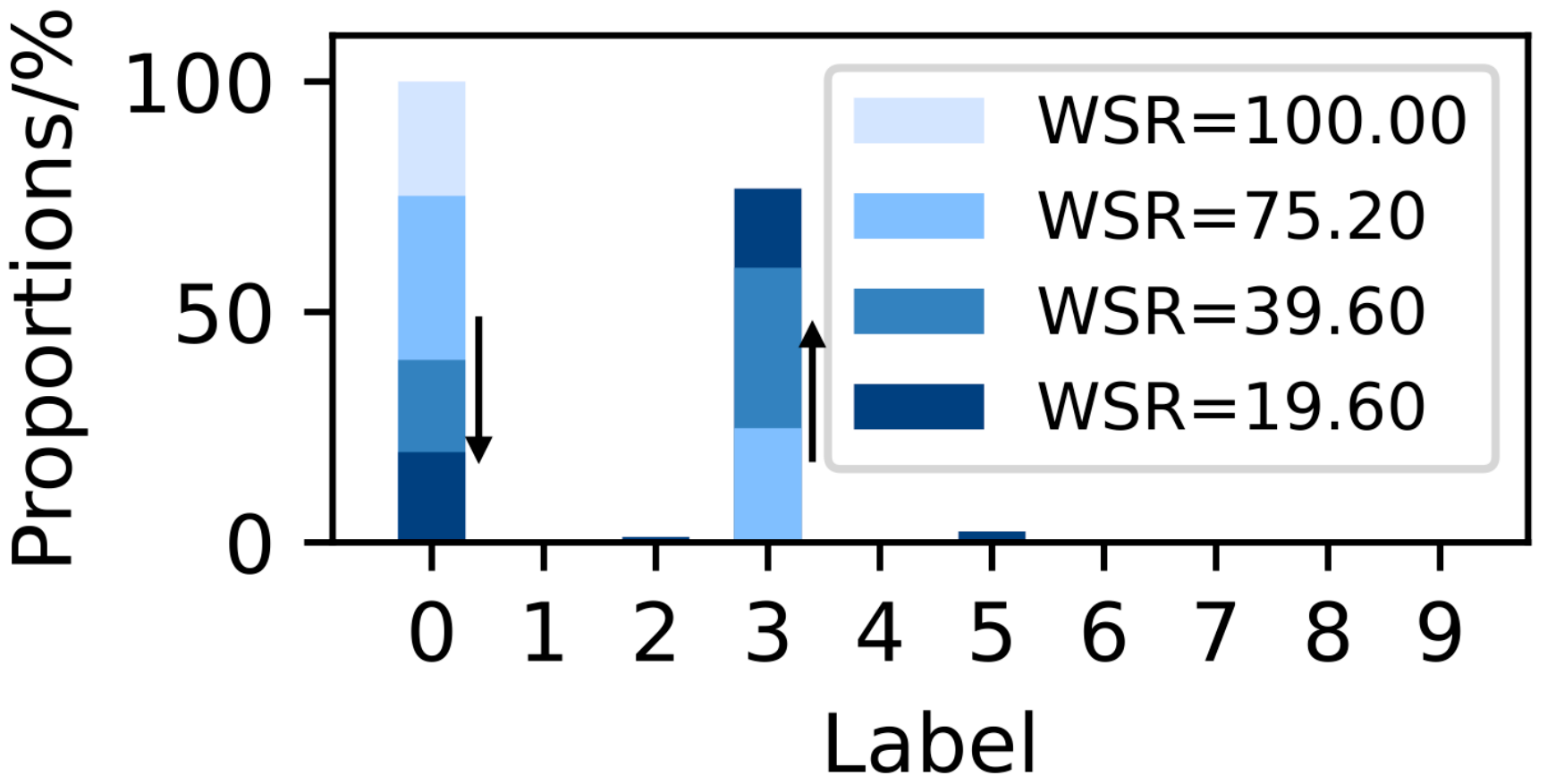}
			\vspace{-0.2cm}
		}\quad
		\subfloat[{\footnotesize No watermark}]{
			\includegraphics[width=0.402\columnwidth]{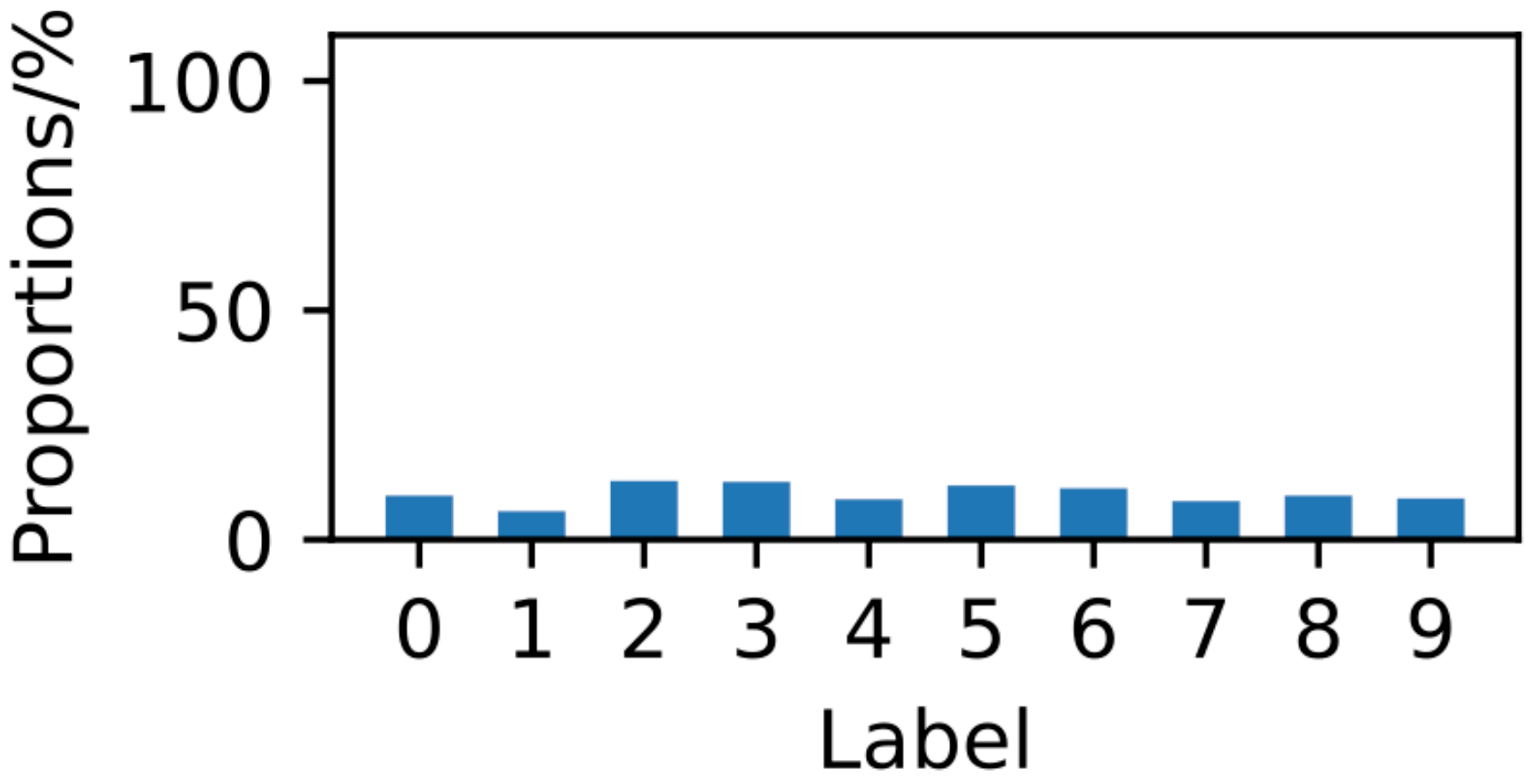}
			\vspace{-0.2cm}
			\label{subfig:cfw_ft_lc}
		}
			\vspace{-0.31cm}
		\caption{Predicted label histograms during WRK attacks.}}
		\vspace{-0.22cm}
\end{figure}

\noindent\textbf{Label Clustering}. We define a \emph{deformation label} to describe new cluster centers that emerge after removal. Figure~\ref{fig:cf_label_cluster} shows label histograms of the watermark task during WRK attacks performed in the section titled \emph{Experiments for WRK}.  After removal, strong clustering persists on both the watermark label ($=0$) and a deformation label ($=3$), due to intra-class representation clustering. Thus, verification with clustering is conducted to evaluate whether the watermark and deformation labels exhibit clustering. The deformation label is not always identical but is predictable and related to the watermark label rather than the watermark samples, which are discussed in Appendix~B~\cite{xiao2025class}.

%% file: experiments.tex
\vspace{-0.10cm}
\section{Experiments for Class-feature Watermarks}
\label{sec:experiment}
\vspace{-0.04cm}
\begin{table*}
	\scriptsize
	\renewcommand{\arraystretch}{0.97}
	\centering
	\begin{tabular}{|>{\raggedleft\arraybackslash}p{0.86cm}>{\raggedleft\arraybackslash}p{1.01cm}| >{\raggedleft\arraybackslash}p{0.86cm}>{\raggedleft\arraybackslash}p{0.87cm} |>{\raggedleft\arraybackslash}p{0.85cm}>{\raggedleft\arraybackslash}p{1.02cm}| >{\centering}p{1.085cm}|>{\raggedleft\arraybackslash}p{0.86cm} >{\raggedleft\arraybackslash}p{0.86cm}>{\raggedleft\arraybackslash}p{1.06cm} |>{\raggedleft\arraybackslash}p{0.86cm}>{\raggedleft\arraybackslash}p{0.9cm} >{\raggedleft\arraybackslash}p{1.06cm}|}
		\hline
	     \multicolumn{2}{|c|}{\textbf{Non-watermark}}&\multicolumn{4}{c|}{\textbf{Victim Model}}&\multicolumn{7}{c|}{\textbf{Copy Model}}\\ \hline
		\multicolumn{2}{|c|}{\textbf{No Removal}}&\multicolumn{2}{c|}{\textbf{No Removal}}&\multicolumn{2}{c|}{\textbf{WRK Removal}}&\multirow{2}{*}{\textbf{MEA}}&\multicolumn{3}{c|}{\textbf{No Removal}}&\multicolumn{3}{c|}{\textbf{WRK Removal}}\\ \cline{1- 6}\cline{8-13}
		\makecell{ACC}&$\text{WSR}_{\text{LC}}$&\makecell{ACC}&$\text{WSR}_{\text{LC}}$&\makecell{ACC}&$\text{WSR}_{\text{LC}}$&&\makecell{ACC}&\makecell{FID}&$\text{WSR}_{\text{LC}}$&\makecell{ACC}&\makecell{FID}&$\text{WSR}_{\text{LC}}$\\ \hline
				\rowcolor{gray!8}
		\multicolumn{13}{|c|}{CIFAR-10}\\ \hline	\multirow{2}{*}{93.55$\pm$0.19;}&\multirow{2}{*}{12.00$\pm$2.67}&\multirow{2}{*}{93.26$\pm$0.12;}&\multirow{2}{*}{100.00$\pm$0}&\multirow{2}{*}{91.95$\pm$0.12;}&\multirow{2}{*}{96.67$\pm$0.67}&MExMI	&89.29$\pm$1.12;&92.70$\pm$1.06;&94.00$\pm$1.33&88.94$\pm$0.95;&90.13$\pm$1.12;&79.33$\pm$2.67\\ \cline{7-13}
        &	&	&	&	&	&ActiveThief	&85.89{$\pm$1.18};	&88.34{$\pm$1.18};	&88.00{$\pm$0.67}	&87.26{$\pm$1.21};	&89.27{$\pm$1.24};	&74.67{$\pm$2.00}
		\\ \hline
				\rowcolor{gray!8}
		\multicolumn{13}{|c|}{CIFAR-20} \\ \hline	\multirow{2}{*}{81.61$\pm$0.35;}&\multirow{2}{*}{6.67$\pm$0.67}&\multirow{2}{*}{81.26$\pm$0.32;}&\multirow{2}{*}{100.00$\pm$0}&\multirow{2}{*}{80.54$\pm$0.21;}&\multirow{2}{*}{96.67$\pm$0.67}&MExMI	&80.64$\pm$1.24;&82.35$\pm$1.36;&85.33$\pm$1.33&80.18$\pm$1.27;&81.97$\pm$1.42;&70.67$\pm$2.67\\ \cline{7-13}
			&	&	&	&	&	&ActiveThief&71.41$\pm$1.33;&77.47$\pm$1.48;&81.33$\pm$1.92&72.15$\pm$1.36;&77.30$\pm$1.54;&64.67$\pm$1.33
		\\ \hline
				\rowcolor{gray!8}
		\multicolumn{13}{|c|}{Imagenette}\\ \hline	\multirow{2}{*}{88.12$\pm$0.28;}&\multirow{2}{*}{11.35$\pm$0.71}&\multirow{2}{*}{85.42$\pm$0.55;}&\multirow{2}{*}{100.00$\pm$0}&\multirow{2}{*}{84.35$\pm$0.42;}&\multirow{2}{*}{95.33$\pm$3.56}&MExMI	&86.04$\pm$1.39;&83.92$\pm$0.66;&86.67$\pm$2.00&84.35$\pm$0.92;&82.93$\pm$0.72;&76.67$\pm$1.23\\ \cline{7-13}
			&	&	&	&	&	&ActiveThief&85.32$\pm$1.14;&82.51$\pm$0.78;&83.33$\pm$1.33&84.49$\pm$0.88;&81.92$\pm$0.84;&73.33$\pm$1.33
		\\ \hline
				\rowcolor{gray!8}
		\multicolumn{13}{|c|}{DBPedia}\\ \hline
		\multirow{2}{*}{98.17$\pm$0.09;}&\multirow{2}{*}{15.39$\pm$0.75}&\multirow{2}{*}{98.03$\pm$0.09;}&\multirow{2}{*}{100.00$\pm$0}&\multirow{2}{*}{97.85$\pm$0.21;}&\multirow{2}{*}{94.92$\pm$0.45}&MExMI&95.15$\pm$0.71;&96.62$\pm$0.96;&97.33$\pm$0.38&94.83$\pm$0.54;&95.76$\pm$1.02;&88.72$\pm$0.42\\  \cline{7-13}
		&&&&&&ActiveThief&91.35$\pm$0.87;&92.82$\pm$1.08;&94.32$\pm$0.57&91.51$\pm$0.63;&93.05$\pm$1.14;&80.64$\pm$0.61
		\\ \hline
				\rowcolor{gray!8}
		\multicolumn{13}{|c|}{Speech Commands}\\ \hline
		\multirow{2}{*}{97.36$\pm$0.12;}&\multirow{2}{*}{8.66$\pm$0.47}&\multirow{2}{*}{97.02$\pm$0.14;}&\multirow{2}{*}{100.00$\pm$0}&\multirow{2}{*}{96.03$\pm$0.24;}&\multirow{2}{*}{95.33$\pm$0.27}&MExMI&96.46$\pm$0.66;&97.82$\pm$1.26;&95.00$\pm$0.73&95.96$\pm$0.69;&96.79$\pm$1.32;&82.12$\pm$0.87\\ \cline{7-13}
		&&&&&&ActiveThief&96.33$\pm$0.72;&97.73$\pm$1.38;&94.00$\pm$0.99&94.61$\pm$0.75;&95.20$\pm$1.44;&79.35$\pm$1.11
		\\ \hline
	\end{tabular}
	\vspace{-0.305cm}
	\caption{Performance of Class-feature Watermark (CFW) Against WRK Removal}
		\label{Table:overall_evaluation}
	\vspace{-0.245cm}
\end{table*}

\begin{table*}
	\scriptsize
	\renewcommand{\arraystretch}{0.954}
	\centering
	\begin{tabular}{>{\centering}p{1.8cm}| >{\raggedleft\arraybackslash}p{0.89cm} >{\raggedleft\arraybackslash}p{1.25cm} >{\raggedleft\arraybackslash}p{0.89cm}>{\raggedleft\arraybackslash}p{0.89cm}>{\raggedleft\arraybackslash}p{0.89cm}>{\raggedleft\arraybackslash}p{1.2cm}|>{\raggedleft\arraybackslash}p{0.9cm}>{\raggedleft\arraybackslash}p{0.9cm}>{\raggedleft\arraybackslash}p{0.94cm}>{\raggedleft\arraybackslash}p{0.94cm}>{\raggedleft\arraybackslash}p{1.18cm}}
		\hline
		\textbf{Watermarks}&\multicolumn{6}{c|}{\textbf{Victim Model}}&\multicolumn{5}{c}{\textbf{Copy Model}}\\ \hline
		\textbf{Metrics}&\makecell[c]{$\mathcal{RE}$$\uparrow$}&\makecell[c]{$\text{CD}^2$$\downarrow$}&\makecell[c]{ACC}&\makecell[c]{WSR}&\makecell[c]{$\text{WSR}_\text{LC}$}&\makecell[c]{Var($\times10^2$)}&\makecell[c]{ACC}&\makecell[c]{FID}&\makecell[c]{WSR}&\makecell[c]{$\text{WSR}_\text{LC}$}&\makecell[c]{Var($\times10^2$)}
		\\ \hline
		w/o $\text{CD}^2$, RepS  &0.19$\pm$0.06&3.98$\pm$0.25&93.70$\pm$0.15&100.00$\pm$0.0&100.00$\pm$0.0&2.59$\pm$0.44&89.55$\pm$1.23&92.54$\pm$1.13&57.33$\pm$2.00&70.00$\pm$2.67&10.30$\pm$3.25
		\\ \hline
		w/ SNNL only &0.85$\pm$0.05&2.51$\pm$0.18&90.13$\pm$0.22&99.33$\pm$0.15&100.00$\pm$0.0&4.14$\pm$1.21&87.18$\pm$1.14&90.49$\pm$0.97&73.33$\pm$2.00&77.33$\pm$2.67&11.12$\pm$3.78
		\\ \hline
		w/ RepS only  &0.92$\pm$0.04&1.78$\pm$0.15&93.31$\pm$0.12&100.00$\pm$0.0&100.00$\pm$0.0&1.92$\pm$0.64&89.85$\pm$1.18&92.85$\pm$1.28&92.00$\pm$1.33&94.00$\pm$1.33&8.17$\pm$2.84
		\\ \hline
		w/ $\text{CD}^2$ only &0.04$\pm$0.03&0.059$\pm$0.008&93.40$\pm$0.14&100.00$\pm$0.0&100.00$\pm$0.0&1.35$\pm$1.02&89.85$\pm$1.21&92.11$\pm$1.32&68.67$\pm$2.00&84.00$\pm$2.67&2.47$\pm$1.01
		\\ \hline
		CFW&0.75$\pm$0.06&0.081$\pm$0.010&93.26$\pm$0.12&100.00$\pm$0.0&100.00$\pm$0.0&1.68$\pm$0.52&89.29$\pm$1.12&92.70$\pm$1.06&91.33$\pm$1.33&94.00$\pm$1.33&2.74$\pm$0.90
		\\ \hline
	\end{tabular}
	
	\hspace{-10.8cm} Arrows represent the trend toward better watermark performance.
		\vspace{-0.305cm}
		\caption{Evaluation Results of Class-feature Watermark (CFW) Variants}
			\label{Table:cf_variants}
	\vspace{-0.41cm}
\end{table*}

\subsection{Overview of Experimental Setup}
\vspace{-0.14cm}
\label{subsec:ex_setup}
We evaluate five tasks spanning three domains: ResNet-18~\cite{he2016deep} trained on \textbf{image} datasets (CIFAR-10, CIFAR-20~\cite{krizhevsky2009learning}), ResNet-50~\cite{he2016deep} trained on ImageNette~\cite{imagenette},  DPCNN~\cite{johnson2017deep} with BERT embeddings trained on a \textbf{text} dataset (DBPedia~\cite{auer2007dbpedia}), and VGG19-BN~\cite{simonyan2014very} trained on an \textbf{audio} dataset (Speech Commands~\cite{warden2018speech}). The CFW dataset is built using multi-class OOD samples assigned to a single label, with size constrained to $0.2\%$--$0.3\%$ of the domain dataset. The watermark samples and MEA query pool are strictly disjoint, with no overlap in classes or distributions. For clustering-based verification, we introduce two metrics: intra-class variance (Var) measuring representation compactness, and label clustering ($\text{WSR}_{\text{LC}}$), defined as the sum of WSRs over watermark and deformation labels. Detailed CFW settings, MEA setups, and metric definitions are provided in Appendix~D~\cite{xiao2025class}.
\vspace{-0.22cm}
\subsection{Overall Evaluation of CFW}
\label{subsec:ex_cfw}
\vspace{-0.115cm}
We evaluate Class-Feature Watermarks (CFW) against the six properties defined in the ownership game. Table~\ref{Table:overall_evaluation} summarizes performance under WRK, focusing on utility preservation (\textbf{Prop.1}), MEA transferability (\textbf{Prop.2}), correctness (\textbf{Prop.3}), and resilience on both victim and copy models (\textbf{Prop.4}, \textbf{Prop.5}). Besides, Appendix~E~\cite{xiao2025class} further evaluates: (1) resilience comparison between CFW and baseline watermarks using WSR and $\text{WSR}_{\text{LC}}$ metrics, (2) resilience under additional removal attacks, (3) impacts of $\text{CD}^2$ and RepS optimization, (4) generalization across architectures, and (5) stealthiness (\textbf{Prop.6}) through anomaly detection analysis.

\noindent\textbf{Results}. 
CFW achieves near-perfect label clustering rates ($\text{WSR}_{\text{LC}} = 100\%$) on watermarked victim models, enabling high-confidence verification. The watermark transfers consistently to copy models extracted via MEA, with $\text{WSR}_{\text{LC}}$ exceeding $81.33\%$ before removal, which attributes to the RepS optimization.  Additionally, model utility degradation remains bounded at $\leq 0.4\%$ except for ImageNette, outperforming existing black-box watermarks which cause $>1.5\%$ degradation (Table~\ref{Table:compare_sta}). On ImageNette, CFW leads to a $2\%$ utility drop because high-resolution inputs amplify the impact of OOD entanglement on the domain task, but its utility remains comparable to benchmarks. Lastly, CFW demonstrates exceptional resilience against removal attacks. Victim models maintain $\text{WSR}_{\text{LC}} \geq 94.67\%$ under WRK attacks, while copy models preserve $\text{WSR}_{\text{LC}} \geq 70.15\%$ despite the combined distortion of MEA and WRK. This robustness stems from two factors: CFW resists WRK’s \emph{Boundary Reshaping} by removing vulnerable decision boundaries, and withstands WRK’s \emph{Feature Shifting} with clustering-based verification, since \emph{Feature Shifting} has a trivial effect on disrupting the representation clustering.
\vspace{-0.25cm}
\subsection{Evaluation on CFW Variants} 
\label{subsec:variant}
\begin{figure}
	\centering{
		\hspace{-0.25cm}\subfloat[\footnotesize WSR]{
			\label{subfig:cfw_d}
			\includegraphics[width=0.318\columnwidth]{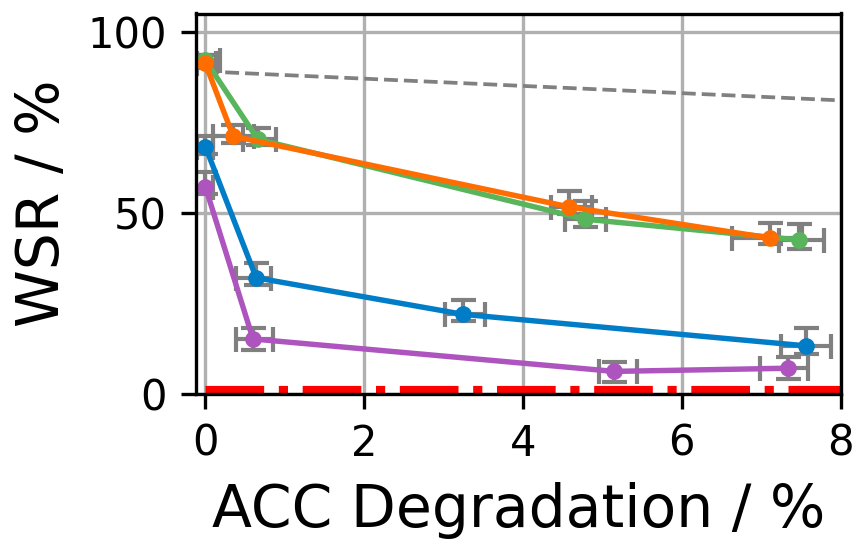}
		}\hspace{-0.12cm}
		\subfloat[{\footnotesize $\text{WSR}_\text{LC}$}]{
			\label{subfig:cfw_e}
			\includegraphics[width=0.318\columnwidth]{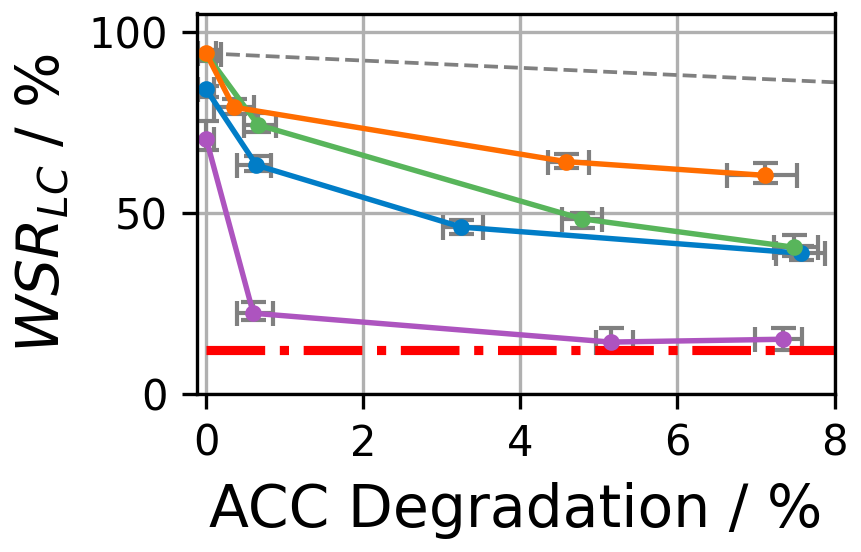}
		}\hspace{-0.12cm}
		\subfloat[{\footnotesize Var}]{
			\label{subfig:cfw_f}
			\includegraphics[width=0.318\columnwidth]{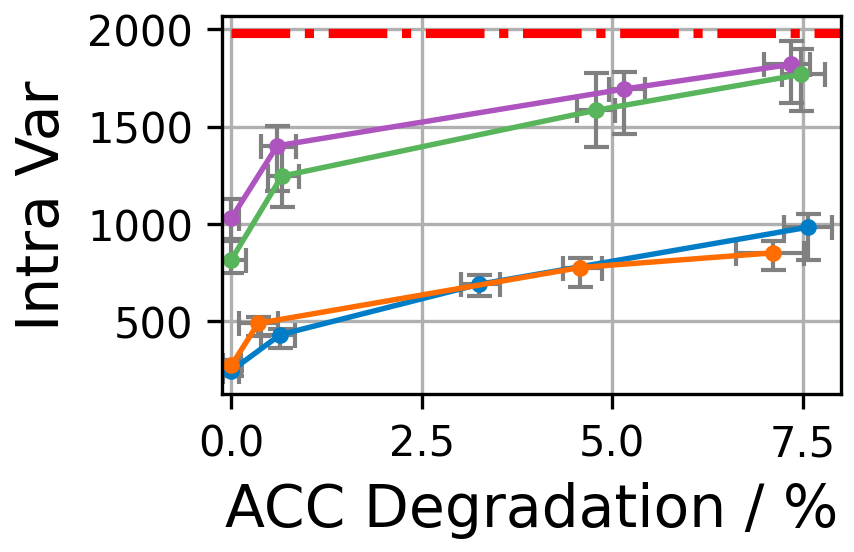}
		}\hspace{-0.1cm}\\
		\includegraphics[width=1.0\columnwidth]{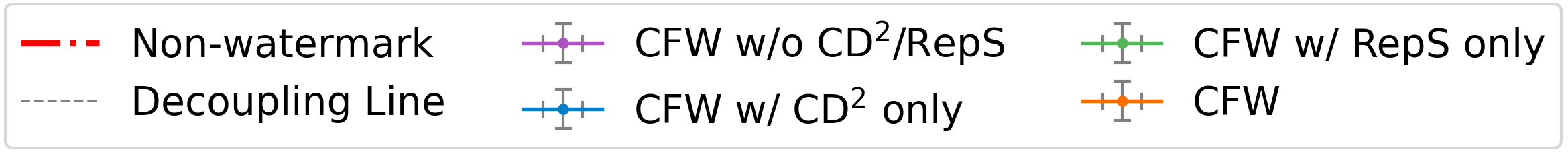}
		\vspace{-0.45cm}
		\caption{Watermark decoupling curves for CFW on extracted copy models. Vertical lines show the error bars. Appendix~E.3 presents corresponding curves on victim models.}
		\label{fig:cfw_decoupling}
	}
	\vspace{-0.445cm}
\end{figure}

\vspace{-0.13cm}
We ablate the two CFW optimization components: RepS and $\text{CD}^2$, and compare RepS with the SNNL~\cite{kornblith2019similarity} algorithm for representation entanglement. The experiments are conduted on CIFAR-10 and further evaluate representation entanglement ($\mathcal{RE}$, Equation~\ref{eq:o}) and pairwise distance projections on distortion ($\text{CD}^2$, Equation~\ref{eq:loss_cd2_final_main}) metrics. Table~\ref{Table:cf_variants} presents quantitative results and Figure~\ref{fig:cfw_decoupling} shows watermark decoupling curves.

\noindent\textbf{Results}. RepS and SNNL both enhance $\mathcal{RE}$ ($0.92$ and $0.85$ vs. baseline $0.19$), but SNNL degrades accuracy by $>3\%$ while delivering inferior transferability. In contrast, RepS maintains model utility ($\Delta$ACC $<0.4\%$) while achieving higher $\mathcal{RE}$ and copy $\text{WSR}_\text{LC}$ ($94.00\%$ vs. SNNL's $77.33\%$). Then, $\text{CD}^2$ optimization reduces $\text{CD}^2$ loss by $30\times$ (to $8.10\times10^{-2}$) and constrains copy model variance to $2.74\times10^2$, which is comparable to victim models ($1.68\times10^2$) and $3.8\times$ lower than the baseline. This directly enhances $\text{WSR}_\text{LC}$ from $70.00\%$ to $94.00\%$ in the full CFW implementation. Finally, Figure~\ref{fig:cfw_decoupling} demonstrates that $\text{WSR}_\text{LC}$ consistently provides clearer evidence than WSR. While RepS only improves WSR resilience, it fails to maintain intra variance clustering. $\text{CD}^2$ optimization enables both robust clustering stability and attack resilience, with $\text{CD}^2$ only variants maintaining label and representation clustering even when WSR drops below $20\%$.

%% file: related_works.tex
\vspace{-0.205cm}
\section{Related Works}
\vspace{-0.105cm}
\label{sec:related_works}
\noindent\textbf{Black-box Watermarks and Their Resilience to Removal.} 
Model watermarking is a promising technique for verifying ownership of machine learning models. While early work showed MEAs could strip watermarks~\cite{lukas2022sok}, recent breakthroughs demonstrate that black-box watermarks can survive extraction attacks by leveraging representation entanglement (RE)~\cite{jia2021entangled,lv2024mea}. 
However, their resilience against consecutive MEA and watermark removal attacks remains unevaluated.

Existing removal methods exploit three decoupling strategies: \textbf{reversion-based removal}, which reconstructs and erases watermark triggers (NC~\cite{wang2019neural}, I-BAU~\cite{zeng2021adversarial}, DTI-DBF~\cite{xu2023towards}, Aiken~\cite{aiken2021neural}); \textbf{neuron pruning}, which eliminates neurons associated with watermark tasks (FP~\cite{liu2018fine}, CLP~\cite{zheng2022data}); and \textbf{learning-induced forgetting}, which removes non-domain features via fine-tuning (NAD~\cite{li2021neural}, SEAM~\cite{zhu2023selective}, FST~\cite{min2023towards}). 
Watermarks with weak entanglement, such as RS~\cite{bansal2022certified} and MBW~\cite{lukas2022sok}, are vulnerable to all three strategies. Strongly entangled watermarks like EWE~\cite{jia2021entangled} and MEA-D~\cite{lv2024mea} resist most removal attacks but remain vulnerable to DTI-DBF and FST. On the other hand, the two removal methods show limited effectiveness against entangled Blend-type watermarks.


%% file: appendix.tex
\section{Appendix}
\section{A. Proof}
\subsection{A.1. Proof of Model Extraction Attack (MEA) Failure Condition in Theorem~\ref{the:ortho}}
\label{app:ortho_proof}
\begin{proof} Assuming $X_q^T$ has a pseudoinverse $(X_q^T)^{-1}$, the copy model's parameters $\theta_{mea}$ estimated by queries are
	\begin{equation}
		\footnotesize
		\theta_{mea} = Y_q^T (X_q^T)^{-1}. 
	\end{equation}
	The performance of $\theta_{mea}$ on a domain sample $\textbf{x}^T\in X$ is
	\begin{equation}
		\label{eq:y_mea}
		\footnotesize
		\textbf{y}_{mea}^T=\theta_{mea} \textbf{x}^T = Y_q^T (X_q^T)^{-1} \textbf{x}^T.
	\end{equation}
	The right term in Equation~\ref{eq:y_mea} performs a weighted sum of the columns of $Y_q^T$, \emph{i.e.},
	\begin{equation}
		\footnotesize
		\textbf{y}_{mea}^T= [\textbf{y}_{q1}^T,\dots] [w_1, \dots]^T=\sum w_i\textbf{y}_{qi}^T,
	\end{equation}
	where $\textbf{y}_{qi}^T$ is the $i$-th column in $Y_q^T$ and $w_i$ is a weight scalar. Since any $\textbf{y}^T \in Y^T$ is orthogonal to all $\textbf{y}_q^T \in Y_q^T$, it cannot be equal to any $ \textbf{y}^T$, \emph{i.e.}, $\textbf{y}_{mea}^T \neq \textbf{y}$.
\end{proof}

\subsection{A.2. Proof of Lower Bound on NTK Cross-Kernel Norm by Representation Entanglement (RE) in Theorem~\ref{theo:relation_re_ntk}}
\label{app:grad_theory_prove}
\noindent\emph{Proof}. 
Following the neural tangent kernel (NTK) theory~\cite{bennani2020generalisation},  we assume the model $F:\mathbb{R}^d\rightarrow\mathbb{R}$. $\|\phi(X_w)\phi(X)^\top\|_2$ can be expressed at the layer level:
\begin{equation}
	\label{eq:layer_grad}
	\footnotesize
	\|\phi(X_w)\phi(X)^\top\|_2=\|\sum\nolimits_{l\in{L}}\nabla_{\theta^l}{F(X_w)}\cdot\nabla_{\theta^l}{F(X)}^\top\|_2.
\end{equation}
Besides, $\nabla_{\theta^l}{F(\cdot)}$ can be calculated by the following equation:
\begin{equation}
	\footnotesize
	\nabla_{\theta^l}{F(\cdot)}= \delta_l(\cdot)F_{\theta^{l-1}}(\cdot),
\end{equation}
where $\delta_l(X_w)=\text{Diag}(\sigma^\prime(z_l))\cdot \theta^{{l+1}\top}\cdot\delta_{l+1}$. For the output layer $L$, $\delta_L=I$. $\sigma$ is the activation function. $z_l$ denotes the output of the $l$-th linear layer, \emph{i.e.}, $F_{\theta^l}(X)=\sigma(z_l)$. $\sigma^\prime$ is the derivative of the activation function. For ReLU, $\sigma^\prime$ consists of 0 or 1.

Assuming that the feature map $\phi(\cdot)$ used in computing RE is defined as the mean vector. Then, Equation~\ref{eq:layer_grad} can be further written as
\begin{equation}
	\footnotesize
	\label{eq:deform_grad}
	\begin{aligned}
		\|&\phi(X_w)\phi(X)^\top\|_2\\
		&=\|\sum_{\mathbf{x}_w\in X_w, }\sum_{\mathbf{x}\in X}\sum_{l\in\{L\}}\delta_l(\mathbf{x}_w)\underbrace{F_{\theta^l}(\mathbf{x}_w)F_{\theta^l}(\mathbf{x})^\top}_{\in \mathbf{R}^{1\times1}}\delta_l(\mathbf{x})^\top\|_2\\
		&=\sum_{\mathbf{x}_w\in X_w}\sum_{\mathbf{x}\in X}|F_{\theta^l}(\mathbf{x}_w)F_{\theta^l}(\mathbf{x})^\top|\underbrace{\|\sum_{l\in\{L\}}\delta_l(\mathbf{x}_w)\delta_l(\mathbf{x})^\top\|_2}_{\gamma(\mathbf{x}_w, \mathbf{x})}\\
	\end{aligned}
\end{equation}
Let $\gamma_{\text{min}} = \min_{\mathbf{x}_w \in X_w,\, \mathbf{x} \in X} \gamma(\mathbf{x}_w, \mathbf{x})$. We note that $\gamma_{\text{min}} \ge 1$ always holds, as formally stated in Theorem~\ref{theo:gamma_min}, which will be proved later. Then from Equation~\ref{eq:deform_grad}, we can derive:
\begin{equation}
	\footnotesize
	\begin{aligned}
		\|\phi(X_w)&\phi(X)^\top\|_2\ge \sum_{\mathbf{x}_w\in X_w}\sum_{\mathbf{x}\in X}\gamma_{\text{min}}\|F_{\theta^l}(\mathbf{x}_w)F_{\theta^l}(\mathbf{x})^\top\|\\
		&\ge \left\| \left(\sum_{\mathbf{x}_w\in X_w}F_{\theta^l}(\mathbf{x}_w)\right)\left(\sum_{\mathbf{x}\in X}F_{\theta^l}(\mathbf{x})\right)^\top 
		\right\|_2\\
		&= \underbrace{N_wN}_{\Gamma}\mathcal{RE}(F_\theta; X_w, X)
	\end{aligned}
\end{equation}
Therefore, let $\Gamma = N_w N$, and we can deduct that:
\begin{equation}
	\footnotesize
	\|\phi(X_w)\phi(X)^\top\|_2\ge\Gamma\cdot\mathcal{RE}(F_\theta; X_w, X)
\end{equation} 

\hfill $\qed$

\begin{theorem}
	\label{theo:gamma_min}
	For any inputs $\mathbf{x}_w \in X_w$, $\mathbf{x} \in X$,
	\begin{equation}
		\footnotesize
		\gamma_{\min} \geq 1,\quad \text{where } \gamma(\mathbf{x}_w, \mathbf{x}) = \left\| \sum_{l=1}^L \delta_l(\mathbf{x}_w)\delta_l(\mathbf{x})^\top \right\|_2
	\end{equation}
\end{theorem}

\noindent\emph{Proof}. 
Each layer contributes a positive semi-definite (PSD) matrix:
\begin{equation}
	\footnotesize
	\delta_l(\mathbf{x}_w)\delta_l(\mathbf{x})^\top \succeq 0 \quad \forall l \in \{1,...,L\},
\end{equation}
where the PSD property follows from the matrix outer product structure. The output layer ($l=L$) contribution is:
\begin{equation}
	\delta_L(\mathbf{x}_w)\delta_L(\mathbf{x})^\top = II^\top = I
\end{equation}
As a result, for the complete sum, the PSD property guarantees spectral norm accumulation:
\begin{equation}
	\footnotesize
	\sum_{l=1}^L \delta_l(\mathbf{x}_w)\delta_l(\mathbf{x})^\top \succeq I + \sum_{l=1}^{L-1} \underbrace{\delta_l(\mathbf{x}_w)\delta_l(\mathbf{x})^\top}_{\succeq 0} \succeq I.
\end{equation}
Therefore, the spectral norm of this PSD matrix satisfies
\begin{equation}
	\footnotesize
	\left\| \sum_{l=1}^L \delta_l(\mathbf{x}_w)\delta_l(\mathbf{x})^\top \right\|_2 \geq \text{largest eigenvalue of } I = 1,
\end{equation}
with equality only if all hidden layer contributions vanish. Since gradient signals in practical networks always contain non-zero backward passes through hidden layers, the inequality is strict ($\gamma_{\min} > 1$) for $L > 1$.
\hfill $\qed$

\subsection{A.3. Derivation of the $\text{CD}^2$ Loss in Equation~\ref{eq:loss_cd2_final_main}}
Starting from neural tangent kernel (NTK) theory~\cite{bennani2020generalisation},  which assumes the model $F:\mathbb{R}^d\rightarrow\mathbb{R}$, \emph{e.g.}, F is a binary classifier, we approximate the output of the MEA-extracted copy model $\hat{F}$ by a linearized model around the victim model $F$:
\begin{equation}
	\footnotesize
	\hat{F}(\textbf{x}) = \phi(\mathbf{x})\Delta\theta + F_0(\mathbf{x}), \quad \text{where}\quad \phi(\mathbf{x}) = \nabla_\theta F(\mathbf{x}),
\end{equation}
and $\Delta\theta$ is derived from fine-tuning data $(X_q, Y_q)$:
\begin{equation}
	\footnotesize
	\Delta\theta = \phi(X_q)^\top\left[\phi(X_q)\phi(X_q)^\top + \lambda I\right]^{-1}\tilde{y},
\end{equation}
with $\tilde{y}=F(X_q)-F_0(X_q)$ representing residual outputs. Considering the closeness of query set $X_q$ to the domain dataset $X$, we approximate $X_q$ with $X$.

To ensure pairwise consistency among watermark representations after distortion, we minimize:
\begin{equation}
	\footnotesize
	\label{eq:eq_loss_cd2_appendix}
	L_{\text{CD}^2}=\frac{1}{N^2_w}\sum\nolimits_{\mathbf{x}_i, \mathbf{x}_j \in D_w} |(\phi(\mathbf{x}_j) - \phi(\mathbf{x}_i))\Delta\theta|.
\end{equation}

Applying singular value decomposition (SVD) to domain gradients, we obtain $\phi(X)=U\Sigma V^T$. Thus, Equation~\ref{eq:eq_loss_cd2_appendix} reduces to minimizing the projection onto principal distortion directions represented by columns of $V$:
\begin{equation}
	\scriptsize
	L_{\text{CD}^2}=\frac{1}{N^2_w}\sum\nolimits_{\mathbf{x}_i, \mathbf{x}_j \in D_w} |(\phi(\mathbf{x}_j)-\phi(\mathbf{x}_i))V\Sigma[\Sigma^2+\lambda I]^{-1}U^\top\tilde{y}|.
\end{equation}

For computational simplicity, we further restrict optimization to the output layer $L$, giving the simplified loss:
\begin{equation}
	\label{eq:loss_cd2_simplified_appendix}
	\scriptsize
	L_{\text{CD}^2}=\frac{1}{N^2_w}\sum\nolimits_{\mathbf{x}_i,\mathbf{x}_j \in D_w}| \left( \nabla_{\theta^L} F(\mathbf{x}_j) - \nabla_{\theta^L} F(\textbf{x}_i) \right) V_{\theta^L} |.
\end{equation}

Considering {$\nabla_{\theta^L}F(\cdot)=F_{\theta^{L-1}}(\cdot)$}, the final simplified form becomes:
\begin{equation}
	\scriptsize
	L_{\text{CD}^2} = \frac{1}{N^2_w}\sum\nolimits_{\mathbf{x}_i,\mathbf{x}_j \in D_w} |\left(F_{\theta^{L-1}}(\mathbf{x}_j) - F_{\theta^{L-1}}(\mathbf{x}_i)\right)V_{\theta^L}|.
\end{equation}

Empirical evidence in Appendix~B.2~\cite{xiao2025class} verifies that the direction represented by $V_{\theta^L}$ aligns closely with the principal component directions of representation distortion induced by MEA, validating our theoretical assumptions.

\section{B. Experimental Validation}
\subsection{B.1. Experimental Validation for the Relationship of Representation Entanglement (RE) and MEA Transferability}
\label{app:ex_re}
We empirically validate the relationship between the metric representation entanglement ($\mathcal{RE}$ in Definition~\ref{def:re}) and MEA transferability across a range of black-box watermark and backdoor methods. $\mathcal{RE}$ is simplified by defining $\psi$ as the mean vector. MEA transferability is quantified by the watermark success rate (WSR) in the extracted copy model (see Equation~\ref{eq:wsr}). In addition to the watermark benchmarks EWE~\cite{jia2021entangled}, MEA-Defender~\cite{lv2024mea} and MBW~\cite{kim2023margin}, we include four different types of backdoor attacks to examine the effect of watermark sample types on MEA transferability, including sticker-type BadNet~\cite{gu2019badnets}, Blend~\cite{chen2017targeted}, invisible-type WaNet~\cite{tuan2021wanet}, and composite backdoors~\cite{lin2020composite}. The $\mathcal{RE}$ of backdoor tasks is influenced by their poison rates~\cite{min2023towards}, with lower rates potentially increasing $\mathcal{RE}$. Therefore, we use minimal poison rates for each backdoor attack, as further reduction fails to embed. We adopt ActiveThief~\cite{pal2020activethief} as the MEA algorithm. Detailed experimental settings are described in Appendix~D~\cite{xiao2025class}. 

\begin{figure}
	\centering{
		\includegraphics[width=0.95\columnwidth]{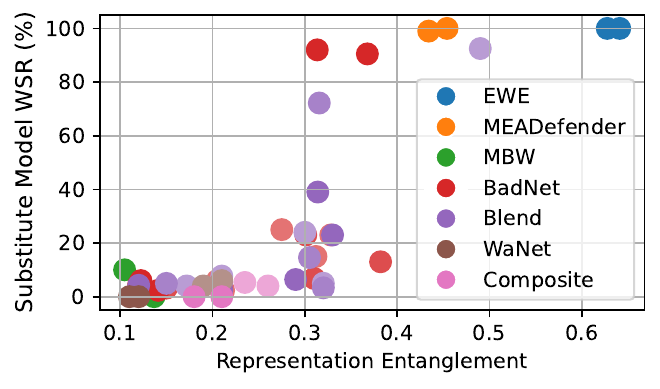}
	}
		\vspace{-0.2cm}
	\caption{WSR of the copy model versus representation entanglement. For each backdoor task, the color intensity indicates the poison rate, with values of $0.001$, $0.002$, and $0.003$ for BadNet and Blend, and $0.05$ and $0.1$ for WaNet and Composite, as lower rates fail to embed.}
	\label{fig:minimal_representation}
	\vspace{-0.35cm}
\end{figure}

\noindent\textbf{Results}. As shown in Figure~\ref{fig:minimal_representation}, there is a clear positive correlation between $\mathcal{RE}$ and the copy model's WSR. Once $\mathcal{RE}$ is lower than $0.3$, WSR drops sharply for backdoor-type tasks. Interestingly, the type of watermark data has a limited impact on WSR, as even sticker-type BadNet achieves high MEA transferability when its $\mathcal{RE}$ is sufficiently large.

\subsection{B.2. Empirical Validation of Distortion Directions Deduced by $\text{CD}^2$ Loss in Equation~\ref{eq:loss_cd2_final_main}}
\label{subsubsec:pc}
In this section, we empirically validate the theoretical objective derived from $\text{CD}^2$ optimization. The term $\left(F_{\theta^{L-1}}(\mathbf{x}_j) - F_{\theta^{L-1}}(\mathbf{x}_i)\right)$ in Equation~\ref{eq:loss_cd2_final_main} can be interpreted as the representation distance between two watermark samples $\mathbf{x}_i$ and $\mathbf{x}_j$, then $V_{\theta^L}$ characterizes \textbf{the distortion direction}s that influence this distance. Intuitively, Equation~\ref{eq:loss_cd2_final_main} minimizes the projection of this distance onto these directions, thereby reducing vulnerability to perturbations.

In Equation~\ref{eq:loss_cd2_final_main}, $V_{\theta^L}$ is defined as the principal component decomposition of the domain dataset's representations. In other words, the distortion directions correspond to the principal directions (PCs) of domain representations. These distortions may result from either MEA or learning-induced removal attacks.

\begin{figure}
	\centering{
		\subfloat[CIFAR10]{
			\includegraphics[width=0.32\columnwidth]{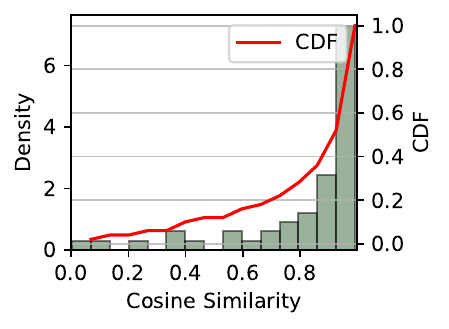}
		}
		\subfloat[CIFAR20]{
			\includegraphics[width=0.32\columnwidth]{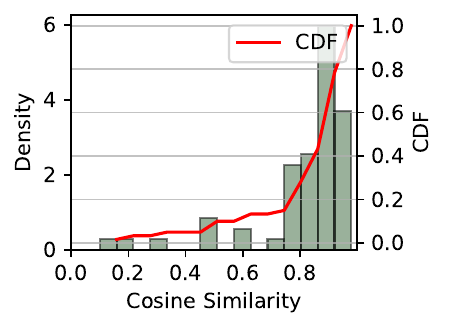}
		}\\		\vspace{-0.2cm}
		\subfloat[DBPedia]{
			\includegraphics[width=0.32\columnwidth]{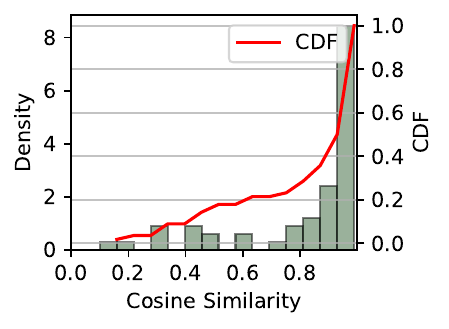}
		}
		\subfloat[Speech Commands]{
			\includegraphics[width=0.32\columnwidth]{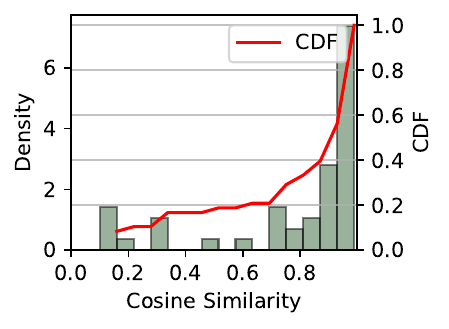}
		}
					\vspace{-0.2cm}
		\caption{Histogram of the domain class PC cosine similarity between original and deformed representations in \textbf{MEA}, with statistics collected over 3-5 trials.}
		\label{fig:pca_mea}
	}
		\vspace{-0.3cm}
\end{figure}

\begin{figure}
	\centering{
		\subfloat[NC]{
			\includegraphics[width=0.32\columnwidth]{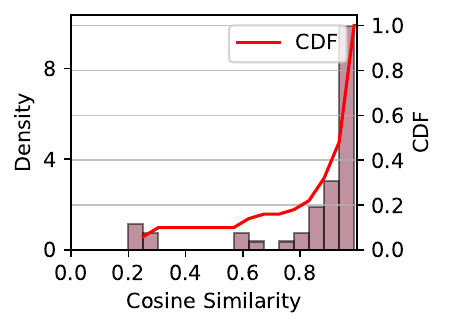}
		}
		\subfloat[I-BAU]{
			\includegraphics[width=0.32\columnwidth]{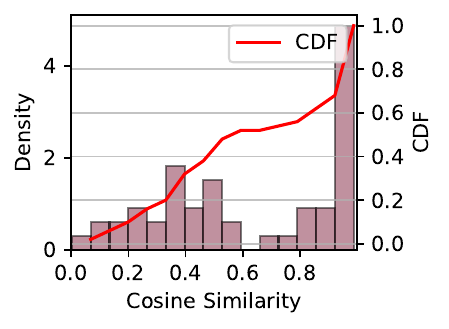}
		}
		\subfloat[NAD]{
			\includegraphics[width=0.32\columnwidth]{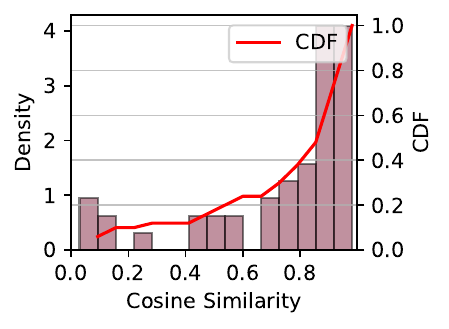}
		}\\
		\subfloat[AT]{
			\includegraphics[width=0.32\columnwidth]{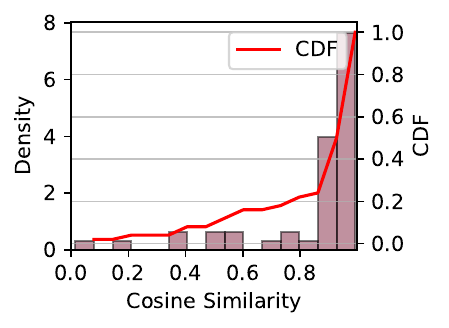}
		}
		\subfloat[WRK]{
			\includegraphics[width=0.32\columnwidth]{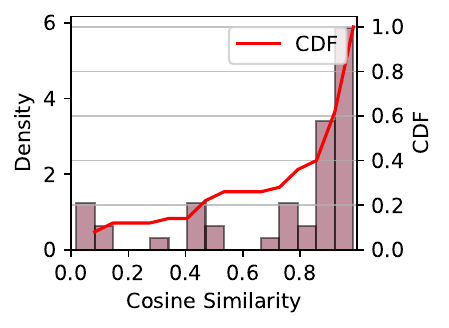}
		}
			\vspace{-0.2cm}
		\caption{Histogram of domain class principal component cosine similarities between original and deformed representations under \textbf{learning-based removal attacks}, averaged over five independent trials.}
		
		\label{fig:pca}
	}
	\vspace{-0.3cm}
\end{figure}

To validate this, we compare the domain PCs of victim models with those of (1) copy models extracted via MEA and (2) models after learning-induced removal. Note that PCs are computed separately for each class to capture fine-grained alignment. The results are shown in Figure~\ref{fig:pca_mea} and Figure~\ref{fig:pca}, respectively. These histograms report the cosine similarity of last-layer PCs before and after distortion. A higher concentration of values near $1$ indicates a more substantial alignment between the updated directions and the PCs of the domain dataset's representations, which supports the deduction in Equation~\ref{eq:loss_cd2_final_main} that these PCs are indeed the distortion directions that affect the watermark data. Our results show that over $80\%$ of cosine similarities exceed $0.6$ across most settings, except for I-BAU. This provides strong empirical support for the $\text{CD}^2$-derived theory that \textbf{the principal directions of domain representations are the dominant distortion directions.
}

\subsection{B.3. Empirical Validation of Predictable Deformation Labels via Principal Component Alignment}
\label{app:deform_label}
In Figure~\ref{fig:cf_label_cluster}, the deformation labels consistently transfer from $0$ to $3$. However, during removal attacks, deformation labels may vary or span multiple classes due to unknown and uneven perturbation strengths across output dimensions.

Nevertheless, we posit that deformation labels are not arbitrary but can be predicted. To substantiate this claim, we first analyze the class-wise interactions during watermark removal. As established in Appendix~B.2~\cite{xiao2025class}, the distortions introduced during removal predominantly align with the principal component (PC) directions of each domain class's representations.

We hypothesize that if the PC directions of a domain class are approximately opposite to those of the watermark-assigned class (computed from the domain dataset), this domain class is more likely to dominate the distortion processes. Consequently, it may absorb the watermark decision boundary, thereby becoming the resulting deformation label. This hypothesis is formalized as follows:

\begin{proposition}[Deformation Label Formation]
	\label{prop:deform}
	Let the following be given:
	\begin{itemize}
		\item $C_d$: A candidate domain class, with its first principal component direction denoted by $\mathbf{v}_d$,
		\item $C_w$: The watermark-assigned class, with its first principal component direction (computed from the domain dataset) denoted by $\mathbf{v}_w$,
		\item $\mathcal{S}(\mathbf{v}_d, \mathbf{v}_w) := \cos(\mathbf{v}_d, \mathbf{v}_w)$: The cosine similarity between $\mathbf{v}_d$ and $\mathbf{v}_w$,
	\end{itemize}
	then $C_d$ emerges as the deformation label under removal perturbation if the following condition holds:
	\begin{equation}
		\footnotesize
		\mathcal{S}(\mathbf{v}_d, \mathbf{v}_w) = \min_{C\sim D}[\mathcal{S}(\mathbf{v}_C, \mathbf{v}_w)],~\&~\mathcal{S}(\mathbf{v}_d, \mathbf{v}_w)<0,
	\end{equation}
	where $D$ is the domain class distribution. This condition indicates geometric opposition in principal subspaces.
\end{proposition}

\begin{figure}
	\centering{
		\includegraphics[width=0.95\columnwidth]{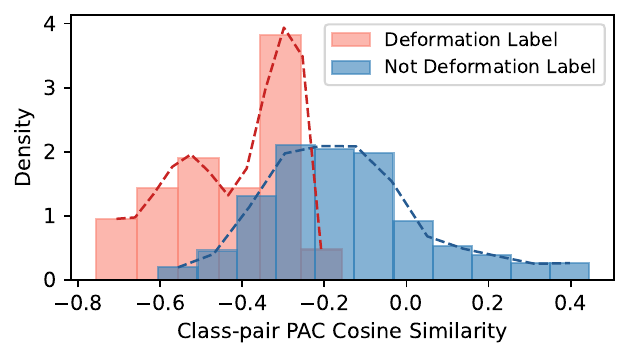}
	}
			\vspace{-0.15cm}
	\caption{\emph{Density} histogram of PC cosine similarity between classes contains watermarks and others. The ratio of deformation labels to non-deformation labels is $1:9$.}
	\label{fig:class_pair_pca}
		\vspace{-0.25cm}
\end{figure}

We validate Proposition~\ref{prop:deform} by empirically analyzing the cosine similarity between the PCs of domain classes assigned watermark tasks and those of their corresponding deformation labels after WRK. Figure~\ref{fig:class_pair_pca} presents the result, which indicates that deformation labels consistently exert opposing stretching, with negative cosine similarity to the watermark label. In contrast, the PCs of non-deformation domain classes exhibit higher orthogonality with the target class assigned to the watermark task. Therefore, to ensure deformation label consistency, the watermark label should be fine-tuned or chosen to have \textbf{only one possible deformation label}, as demonstrated with class $0$ in the experiments in the section titled \emph{Verify CFW with Intra-class Clustering}.

\section{C. Algorithm of Watermark Removal Attack}
\label{app:ahlg_wrk_attack}
Algorithm~\ref{alg:wrk_attack} presents the Watermark Removal Attack (WRK) in algorithmic form.

\begin{algorithm}
	\small
	\renewcommand{\algorithmicrequire}{\textbf{Input:}}
	\renewcommand{\algorithmicensure}{\textbf{Output:}} 
	\caption{Watermark Removal Attack (WRK)}
	\label{alg:wrk_attack}
	\begin{algorithmic}[1]
		\REQUIRE Stolen copy model $F_s$ with parameters $\theta$, 
		domain-relevant subset $D_d$, 
		subset ratio $\rho \in (0,1]$,
		noise magnitudes $\epsilon$,
		\ENSURE Watermark removed model $F_{s^{\text{wrk}}}$
		\STATE Extract final layer weights: $\theta^L_{\text{ini}} \gets \theta^L$
		\STATE Sample subset: $D_d' \gets \textsc{RandomSample}(D_d, \rho)$
		
		\FOR {each $x \in D_d'$}
		\STATE Compute perturbation: $\delta_x \gets \textsc{FGSM}(F_\theta, x, \epsilon)$
		\STATE Generate adversarial example: $\tilde{x} \gets \text{Clip}(x + \delta_x)$
		\STATE Assign random label: $y \sim \mathcal{U}(\{1,...,K\})$
		\STATE Add to the poisoned set: $D_p \gets D_p \cup \{(\tilde{x}, y)\}$
		\ENDFOR
		\STATE Combine datasets: $D_{\text{train}} \gets D_d \cup D_p$
		\STATE Initialize $\theta^L$
		\STATE Fine-tune model: {$F_{s^{\text{wrk}}} \gets \text{argmin}_\theta L_\text{wrk}$ } \hfill $\triangleright$ Equation~\ref{eq:wrk_loss}
		\RETURN $F_{s^{\text{wrk}}}$
	\end{algorithmic}
\end{algorithm}

\section{D. Experimental Setup Details}
\label{app:ex_wrk_setup_full}
\subsection{D.1. WRK Experiment Setup Details} 
\noindent\textbf{Black-box Watermark and Backdoor Benchmarks}. We benchmark four black-box model watermarks against MEAs: \textbf{EWE}~\cite{jia2021entangled}, \textbf{MBW}~\cite{kim2023margin}, \textbf{MEA-Defender (MEA-D)}~\cite{lv2024mea}, and a typical backdoor methods \textbf{Blend}~\cite{chen2017targeted}. Since these watermarking methods are specifically designed for image classification, we conduct experiments on the CIFAR-10~\cite{krizhevsky2009learning} dataset with ResNet18~\cite{he2016deep} and on ImageNette~\cite{imagenette} with ResNet50~\cite{he2016deep}. All experiments follow their original frameworks with optimized settings. In detail, for EWE and MEA-Defender, the watermark datasets comprise $5\%$ and $10\%$ of the training data, respectively. MBW follows its original setups, using 10 watermark samples. Blend adopts a small watermark dataset ratio of $0.3\%$, which contributes to increasing its representation entanglement (RE). 

To assess the broader applicability of WRK, we evaluate its removal performance on other backdoors: \textbf{WaNet}~\cite{tuan2021wanet} and \textbf{composite backdoor}~\cite{lin2020composite}. These backdoor methods are excluded from watermark evaluation since they do not survive model extraction.

\noindent\textbf{Removal-related Setups}. We compare WRK against nine recent removal methods that have official implementations and are not limited by architectural assumptions (\emph{e.g.}, batch normalization). These methods fall into three categories: \textbf{reversion-based} (NC~\cite{wang2019neural}, I-BAU~\cite{zeng2021adversarial}, DTI-DBF~\cite{xu2023towards}), \textbf{neuron pruning} (FP~\cite{liu2018fine}, CLP~\cite{zheng2022data}), and \textbf{learning-induced forgetting} (NAD~\cite{li2021neural}, SEAM~\cite{zhu2023selective}, FST~\cite{min2023towards}). In addition, Adversarial Training (AT)~\cite{kurakin2016adversarial} is included as a baseline.

All removal methods use a domain dataset comprising $5\%$ of the original training set for CIFAR-10 and $10\%$ for ImageNette, and model utility degradation is constrained to within $2\%$ where applicable. For WRK, the noise magnitude $\epsilon$ is set to $1.5_{-0.5}^{+0.5}$ (after normalization) and the coefficient for weight-shifting regularization $\alpha$ is set to $0.05_{-0.02}^{+0.05}$. The optimizer in WRK is set to SGD~\cite{bottou2010large} with learning rate $0.001$ for CIFAR-10 and $0.0003_{-0.0002}^{+0.0002}$ for Imagenette.

\noindent\textbf{Model Extraction Attack (MEA) Settings}. Considering the threat model assumes the adversary prepares a limited number of domain samples for removal attacks, we evaluate two MEA benchmarks: MExMI~\cite{xiao2022mexmi}, which emphasizes domain samples to enhance MEA performance, and ActiveThief~\cite{pal2020activethief}. To ensure meaningful attack results, the query data pool is supplemented with out-of-domain datasets from ImageNet~\cite{deng2009imagenet}. Both methods use a query budget of 25,000 for CIFAR-10 and 5,000 for ImageNette, which are consistent with their original scales.

\noindent\textbf{Metrics}. The calculations for these metrics are as follows:

\noindent\textbf{Accuracy (ACC)}. For the model $F$ and the test dataset $D_t$ of size $N_t$, 
	\vspace{-0.25cm}
\begin{equation}
	\footnotesize
	\text{ACC}(F, D_t)=\frac{100}{N_t}\sum\nolimits_{(\mathbf{x},y)\in D_t}{\mathbf{1}(F(\mathbf{x})=y})\%, 
		\vspace{-0.08cm}
\end{equation}
where $\mathbf{1}(\cdot)$ denotes the indicator function.

\noindent\textbf{Fidelity (FID)}. For the copy model $F_s$ and the victim model $F_v$, 
		\vspace{-0.25cm}
\begin{equation}
	\footnotesize
	\text{FID}(F_s, F_v)=\frac{100}{N_t}\sum\nolimits_{(\mathbf{x},y)\in D_t}{\mathbf{1}(F_s(\mathbf{x})=F_v(\mathbf{x})})\%.
		\vspace{-0.08cm}
\end{equation}	

\noindent\textbf{Watermark Success Rate (WSR)}. For the model $F$ and  the watermark dataset $D_w$ of size $N_w$ and the watermark label $y_\text{w}$,
		\vspace{-0.25cm}
\begin{equation}
	\footnotesize
	\label{eq:wsr}
	\text{WSR}(F, D_w | y_\text{w})=\frac{100}{N_w}\sum\nolimits_{\mathbf{x}\in D_w}{\mathbf{1}(F(\mathbf{x})=y_\text{w})}\%.
		\vspace{-0.08cm}
\end{equation}

\noindent\textbf{Implementation Details}. All results are averaged over five independent runs with different random seeds to stimulate uncertainty, with variation bounds shown as \textbf{maximum} observed deviations. The implementation runs on a Windows workstation equipped with two NVIDIA RTX 4090 GPUs.

\subsection{D.2. CFW Experiment Setup Details} 
\noindent\textbf{Dataset Descriptions}. CIFAR-10 and CIFAR-20 each contain 50,000 training and 10,000 test images (3-channel 32x32 pixels); CIFAR-10 has 10 labels, while CIFAR-20 uses 20 superclasses from CIFAR-100. ImageNette contains higher-resolution images which we resize to 128x128 pixels, comprising 9,469 training instances and 3,925 test samples across 10 categories. DBPedia includes 560,000 samples categorized into 14 classes. Google Speech Commands comprises over 105,000 utterances of 35 words from various speakers, organized into 12 classes.

\noindent\textbf{Class-Feature Watermark (CFW) Settings}. The CFW dataset is created by selecting multiple out-of-domain (OOD) data types and assigning them to the same label. To constrain feature complexity, the watermark dataset is limited to $0.2\%$–$0.3\%$ of the domain dataset, except for Imagenette, which uses $1.5\%$. For CIFAR-10, 150 samples are taken from 10 non-overlapping classes in CIFAR-100. For CIFAR-20, 150 samples are selected from 4 classes in ImageNet~\cite{deng2009imagenet}. For ImageNette, 30 samples are drawn from 10 ImageNet classes that do not overlap with those in ImageNette, even at the superclass level. For DBPedia, 300 CFW samples are LLM-generated to mimic DBPedia’s style while remaining semantically OOD, and for Google Speech Commands, 300 samples are taken from 4 classes within the `unknown' category. The weighting factors $\lambda_1$ is set to $0.5_{-0.3}^{+0.3}$, and $\lambda_2 > 0$ is set to $0.0003_{-0.0002}^{+0.0005}$.

\noindent\textbf{Model Extraction Attack (MEA) Settings}. The MEA query pools of MExMI~\cite{xiao2022mexmi} and ActiveThief~\cite{pal2020activethief} consist of both in-domain and out-of-distribution (OOD) samples. \textbf{Notably, the OOD samples used in MEA are entirely disjoint from the CFW dataset, and even the classes do not overlap}. For instance, in the CIFAR-20 experiment, the adversarial pool consists of the first 100 ImageNet classes, while the CFW data is drawn from classes 530 to 533. Following the WRK experiment setups, the domain subset constitutes 5\% of the training dataset and is also used for removal attacks. For image tasks CIFAR-10 and CIFAR-20, the OOD samples are from ImageNet32~\cite{deng2009imagenet}, with a query budget of 25,000. For ImageNette, OOD samples are selected from ImageNet with a budget of 5,000. For DBPedia, the pool is AG News dataset~\cite{zhang2015character}, with a budget of 50,000, and for Speech Commands, the pool includes classes from version 0.02 absent in version 0.01, with a budget of 100,000. The copy model architectures match the victim models in default experiments. Further, Appendix~E.10~\cite{xiao2025class} reports cross-experiments using ResNet-18, MobileNetV2~\cite{sandler2018mobilenetv2}, and VGG19-BN to evaluate the impact of architectures on CFW. 

\noindent\textbf{Metrics}. Since CFW is verified with clustering, in addition to WSR, we introduce the following two metrics.

\noindent\textbf{Intra-class Variance (Intra Var, Var)}. This metric calculates the mean squared distance from each sample to its class centroid in the t-SNE~\cite{van2008visualizing} reduced representation space (normalized to [-100, 100]). For a model $F$ and the watermark dataset $D_w$, it is computed as:
\begin{equation}
	\label{eq:intra_var}
	\footnotesize
	\text{Var}(F, D_w)=\frac{1}{N_w}\sum\nolimits_{\mathbf{x}\in D_w}{\|\text{t-SNE}(F_{\theta^L}(\mathbf{x}))-\mu\|}_2^2,
\end{equation}
where $\mu$ is the centroid of the reduced representations of watermark tasks, $\mu=\mathbf{E}(\text{t-SNE}(F_{\theta^L}(D_w)))$.

\noindent\textbf{Label Clustering ($\text{WSR}_\text{LC}$)}. This metric evaluates the clustering behavior on the watermark label $y_\text{w}$ and the deformation label $y_\text{deform}$:
\begin{equation}
	\label{eq:l_clustering}
	\footnotesize
	\text{WSR}_\text{LC}=\text{WSR}(F, D_w | y_\text{w}) + \text{WSR}(F, D_w | y_\text{deform}).
\end{equation}

\begin{table*}
	\tiny
	\renewcommand{\arraystretch}{1.12}
	\caption{Performance of WRK and Benchmark Removal Attack on ImageNette}
	\vspace{-0.2cm}
	\label{app:table:removal_attack}
	\centering
	\begin{tabular}{p{2.056cm}p{0.375cm}p{0.385cm} p{0.35cm}p{0.38cm} p{0.38cm}p{0.38cm}p{0.38cm}p{0.38cm} p{0.38cm}p{0.38cm} p{0.38cm}p{0.38cm} p{0.38cm}p{0.38cm} p{0.395cm}p{0.38cm} p{0.395cm}p{0.38cm} p{0.38cm}p{0.405cm}}
		\hline
		\multicolumn{1}{>{\centering}p{0.995cm}|}{\textbf{Removal}}&\multicolumn{2}{>{\centering}p{1.302cm}}{\multirow{2}{*}{\textbf{None}}}
		&\multicolumn{6}{|>{\centering}p{4.15cm}|}{\textbf{Reversion-type Removal}}&\multicolumn{4}{>{\centering}p{2.7cm}|}{\textbf{Neuron Pruning}}&\multicolumn{8}{>{\centering}p{5.45cm}}{\textbf{Learning-induced Forgetting}} \\ \cline{4-21}
		\multicolumn{1}{>{\centering}p{0.995cm}|}{\textbf{Method}}&&&\multicolumn{2}{|>{\centering}p{1.15cm}|}{\textbf{NC}}&\multicolumn{2}{>{\centering}p{1.15cm}|}{\textbf{I-BAU}}&\multicolumn{2}{c|}{\textbf{BTI-DBF}}&\multicolumn{2}{c|}{\textbf{CLP}}&\multicolumn{2}{c|}{\textbf{FP}}&\multicolumn{2}{c|}{\textbf{NAD}}&\multicolumn{2}{c|}{\textbf{SEAM}}&\multicolumn{2}{c|}{\textbf{FST}}&\multicolumn{2}{c}{\textbf{WRK(Ours)}} \\ \hline
		\multicolumn{1}{>{\centering}p{0.995cm}|}{\textbf{Metrics}/\%}&ACC&WSR&ACC&WSR&ACC&WSR&ACC&WSR&ACC&WSR&ACC&WSR&{ACC}&{WSR}&ACC&WSR&ACC&WSR&ACC&WSR\\ \hline
	\end{tabular}
	\begin{tabular}{>{\centering}p{1.025cm}|>{\raggedleft\arraybackslash}p{0.38cm}>{\raggedleft\arraybackslash}p{0.42cm} >{\raggedleft\arraybackslash}p{0.372cm}>{\raggedleft\arraybackslash}p{0.425cm} >{\raggedleft\arraybackslash}p{0.382cm}>{\raggedleft\arraybackslash}p{0.395cm} >{\raggedleft\arraybackslash}p{0.372cm}>{\raggedleft\arraybackslash}p{0.385cm} >{\raggedleft\arraybackslash}p{0.372cm}>{\raggedleft\arraybackslash}p{0.38cm} >{\raggedleft\arraybackslash}p{0.372cm}>{\raggedleft\arraybackslash}p{0.38cm} >{\raggedleft\arraybackslash}p{0.372cm}>{\raggedleft\arraybackslash}p{0.38cm} >{\raggedleft\arraybackslash}p{0.372cm}>{\raggedleft\arraybackslash}p{0.38cm} >{\raggedleft\arraybackslash}p{0.372cm}>{\raggedleft\arraybackslash}p{0.38cm} >{\raggedleft\arraybackslash}p{0.395cm}>{\raggedleft\arraybackslash}p{0.52cm}}
		EWE&86.42\scalebox{0.4}{$\pm$0.38}&99.95\scalebox{0.4}{$\pm$0.05}&84.76\scalebox{0.4}{$\pm$0.42}&99.61\scalebox{0.4}{$\pm$0.15}&86.28\scalebox{0.4}{$\pm$0.35}&99.12\scalebox{0.4}{$\pm$0.18}&85.92\scalebox{0.4}{$\pm$0.42}&92.59\scalebox{0.4}{$\pm$0.66}&85.89\scalebox{0.4}{$\pm$0.38}&98.75\scalebox{0.4}{$\pm$0.25}&84.52\scalebox{0.4}{$\pm$0.42}&98.72\scalebox{0.4}{$\pm$0.22}&84.39\scalebox{0.4}{$\pm$0.49}&98.56\scalebox{0.4}{$\pm$0.34}&83.25\scalebox{0.4}{$\pm$0.52}&99.22\scalebox{0.4}{$\pm$0.13}&84.03\scalebox{0.4}{$\pm$0.63}&69.14\scalebox{0.4}{$\pm$10.3}&84.35\scalebox{0.4}{$\pm$0.37}&\textbf{10.55}\scalebox{0.4}{$\pm$0.74}
		\\
		MBW&76.01\scalebox{0.4}{$\pm$1.13}&100.00\scalebox{0.4}{$\pm$0}&75.53\scalebox{0.4}{$\pm$1.25}&10.00\scalebox{0.4}{$\pm$10.0}&74.85\scalebox{0.4}{$\pm$1.15}&6.00\scalebox{0.4}{$\pm$6.00}&74.21\scalebox{0.4}{$\pm$1.35}&2.00\scalebox{0.4}{$\pm$8.00}&72.03\scalebox{0.4}{$\pm$1.45}&46.00\scalebox{0.4}{$\pm$34.0}&74.26\scalebox{0.4}{$\pm$1.30}&84.00\scalebox{0.4}{$\pm$6.00}&74.14\scalebox{0.4}{$\pm$1.20}&4.00\scalebox{0.4}{$\pm$6.00}&79.15\scalebox{0.4}{$\pm$1.05}&\textbf{0.00}\scalebox{0.4}{$\pm$0.00}&78.85\scalebox{0.4}{$\pm$1.10}&8.00\scalebox{0.4}{$\pm$8.00}&79.39\scalebox{0.4}{$\pm$0.95}&6.00\scalebox{0.4}{$\pm$6.00}
		\\
		MEA-D&74.36\scalebox{0.4}{$\pm$0.75}&79.54\scalebox{0.4}{$\pm$0.66}&73.51\scalebox{0.4}{$\pm$0.82}&70.11\scalebox{0.4}{$\pm$1.24}&73.19\scalebox{0.4}{$\pm$0.82}&33.88\scalebox{0.4}{$\pm$10.6}&73.42\scalebox{0.4}{$\pm$0.79}&14.34\scalebox{0.4}{$\pm$8.94}&72.53\scalebox{0.4}{$\pm$0.92}&77.82\scalebox{0.4}{$\pm$0.74}&72.97\scalebox{0.4}{$\pm$0.44}&78.84\scalebox{0.4}{$\pm$0.59}&73.25\scalebox{0.4}{$\pm$0.66}&24.82\scalebox{0.4}{$\pm$5.38}&73.95\scalebox{0.4}{$\pm$0.63}&16.97\scalebox{0.4}{$\pm$7.13}&72.93\scalebox{0.4}{$\pm$0.86}&36.84\scalebox{0.4}{$\pm$6.46}&74.06\scalebox{0.4}{$\pm$0.53}&\textbf{3.62}\scalebox{0.4}{$\pm$0.81}
		\\
		Blend&82.59\scalebox{0.4}{$\pm$0.65}&90.52\scalebox{0.4}{$\pm$1.25}&N/A&N/A&82.36\scalebox{0.4}{$\pm$0.68}&82.52\scalebox{0.4}{$\pm$1.52}&80.52\scalebox{0.4}{$\pm$0.75}&56.75\scalebox{0.4}{$\pm$9.75}&80.53\scalebox{0.4}{$\pm$0.76}&81.67\scalebox{0.4}{$\pm$1.65}&83.17\scalebox{0.4}{$\pm$0.51}&80.76\scalebox{0.4}{$\pm$1.69}&80.67\scalebox{0.4}{$\pm$0.88}&77.73\scalebox{0.4}{$\pm$1.55}&80.27\scalebox{0.4}{$\pm$0.85}&19.66\scalebox{0.4}{$\pm$6.52}&80.35\scalebox{0.4}{$\pm$0.92}&6.35\scalebox{0.4}{$\pm$1.85}&80.56\scalebox{0.4}{$\pm$0.45}&\textbf{3.53}\scalebox{0.4}{$\pm$0.91}
		\\
		WaNet&87.16\scalebox{0.4}{$\pm$0.45}&99.15\scalebox{0.4}{$\pm$0.25}&85.54\scalebox{0.4}{$\pm$0.50}&43.53\scalebox{0.4}{$\pm$8.35}&87.28\scalebox{0.4}{$\pm$0.40}&\textbf{0.15}\scalebox{0.4}{$\pm$0.16}&85.17\scalebox{0.4}{$\pm$0.55}&0.31\scalebox{0.4}{$\pm$0.31}&86.68\scalebox{0.4}{$\pm$0.34}&99.89\scalebox{0.4}{$\pm$0.15}&85.86\scalebox{0.4}{$\pm$0.60}&4.85\scalebox{0.4}{$\pm$1.24}&85.27\scalebox{0.4}{$\pm$0.65}&0.36\scalebox{0.4}{$\pm$0.36}&84.94\scalebox{0.4}{$\pm$0.74}&0.42\scalebox{0.4}{$\pm$0.42}&85.24\scalebox{0.4}{$\pm$0.75}&0.83\scalebox{0.4}{$\pm$0.83}&86.89\scalebox{0.4}{$\pm$0.30}&0.25\scalebox{0.4}{$\pm$0.25}
		\\
		Composite&86.78\scalebox{0.4}{$\pm$0.50}&94.14\scalebox{0.4}{$\pm$0.85}&84.70\scalebox{0.4}{$\pm$0.55}&82.80\scalebox{0.4}{$\pm$1.23}&85.15\scalebox{0.4}{$\pm$0.45}&73.45\scalebox{0.4}{$\pm$7.65}&83.49\scalebox{0.4}{$\pm$0.68}&11.18\scalebox{0.4}{$\pm$2.05}&86.88\scalebox{0.4}{$\pm$0.39}&92.87\scalebox{0.4}{$\pm$0.95}&84.32\scalebox{0.4}{$\pm$0.65}&98.60\scalebox{0.4}{$\pm$0.35}&84.68\scalebox{0.4}{$\pm$0.78}&49.20\scalebox{0.4}{$\pm$11.8}&86.19\scalebox{0.4}{$\pm$0.35}&3.36\scalebox{0.4}{$\pm$0.95}&84.61\scalebox{0.4}{$\pm$0.80}&\textbf{0.59}\scalebox{0.4}{$\pm$0.35}&86.19\scalebox{0.4}{$\pm$0.25}&5.96\scalebox{0.4}{$\pm$0.85}
		\\ \hline
	\end{tabular}
	
	\hspace{-9.3cm} N/A indicates that NC does not identify any suspicious class and thus does not perform removal.
	\vspace{-0.3cm}
\end{table*}

\section{E. Supplementary Experiments}
\subsection{E.1. WRK Performance on Imagenette}
\label{subsec:ex_wrk_benchmark_removal}
We extend our experiments of WRK to ImageNette with ResNet50. Table~\ref{app:table:compare_sta} confirms that EWE and MEA‑Defender transfer well under MExMI and ActiveThief, whereas MBW does not. Blend is omitted because it never transfers on this dataset.

Table~\ref{app:table:removal_attack} compares WRK with existing removal methods on victim models. In detail, reversion-type removal, neuron pruning and learning-induced forgetting baselines leave high WSRs ($\leq70\%$) on strong watermarks such as EWE. Yet WRK cuts WSRs below $11\%$ for every watermark while keeping accuracy high. WRK also cleanses the backdoors, reducing the success rate of WaNet to $0\%$ WSR and Composite to $5.96\%$. These results confirm that WRK is still effective on a high-resolution image benchmark.

\begin{table}
	\tiny
	\renewcommand{\arraystretch}{1.09}
	\centering
	\caption{Performance of Existing Black-box Watermarks on ImageNette}
	\vspace{-0.25cm}
	\label{app:table:compare_sta}
	\begin{tabular}{>{\centering}p{0.685cm}|p{0.42cm}>{\raggedleft\arraybackslash}p{0.455cm}>{\raggedleft\arraybackslash}p{0.42cm}>{\raggedleft\arraybackslash}p{0.445cm}>{\centering}p{0.715cm}>{\raggedleft\arraybackslash}p{0.43cm}>{\raggedleft\arraybackslash}p{0.43cm}>{\raggedleft\arraybackslash}p{0.53cm}}
		\hline
		\textbf{Method}&\multicolumn{2}{c|}{\textbf{Non-WM}}&\multicolumn{2}{c|}{\textbf{Victim Model}}&\multicolumn{4}{c}{\textbf{Copy Model}}\\ \hline
		\textbf{Metric}/\%&ACC&WSR&ACC&WSR&\textbf{MEA}&{ACC}&{FID}&WSR\\ \hline
	\end{tabular}
	\begin{tabular}{>{\centering}p{0.685cm}|p{0.42cm}>{\raggedleft\arraybackslash}p{0.455cm}>{\raggedleft\arraybackslash}p{0.40cm}>{\raggedleft\arraybackslash}p{0.445cm}>{\centering}p{0.715cm}>{\raggedleft\arraybackslash}p{0.43cm}>{\raggedleft\arraybackslash}p{0.43cm}>{\raggedleft\arraybackslash}p{0.53cm}}
		\multirow{2}{*}{EWE}&\multirow{2}{*}{88.12\scalebox{0.4}{$\pm$0.15}}&\multirow{2}{*}{14.29\scalebox{0.4}{$\pm$3.32}}&\multirow{2}{*}{86.42\scalebox{0.4}{$\pm$0.38}}&\multirow{2}{*}{\hspace{-0.03cm}99.95\scalebox{0.4}{$\pm$0.05}}&MExMI&83.53\scalebox{0.4}{$\pm$0.52}&83.63\scalebox{0.4}{$\pm$0.60}&60.16\scalebox{0.4}{$\pm$2.75}\\
		&&&&&ActiveThief&81.54\scalebox{0.4}{$\pm$0.69}&81.96\scalebox{0.4}{$\pm$0.75}&58.79\scalebox{0.4}{$\pm$2.80}
		\\ \hline
		\multirow{2}{*}{MBW}&\multirow{2}{*}{88.12\scalebox{0.4}{$\pm$0.15}}&\multirow{2}{*}{8.00\scalebox{0.4}{$\pm$8.00}}&\multirow{2}{*}{76.01\scalebox{0.4}{$\pm$1.13}}&\multirow{2}{*}{\hspace{-0.04cm}100.00\scalebox{0.4}{$\pm$0}}&MExMI&72.85\scalebox{0.4}{$\pm$0.75}&83.05\scalebox{0.4}{$\pm$0.79}&0.00\scalebox{0.4}{$\pm$0.00}\\
		&&&&&ActiveThief&70.16\scalebox{0.4}{$\pm$0.82}&79.55\scalebox{0.4}{$\pm$0.95}&0.00\scalebox{0.4}{$\pm$0.00}
		\\ \hline
		\multirow{2}{*}{\makecell[c]{MEA-D}}&\multirow{2}{*}{88.12\scalebox{0.4}{$\pm$0.15}}&\multirow{2}{*}{~1.73\scalebox{0.4}{$\pm$0.48}}&\multirow{2}{*}{74.36\scalebox{0.4}{$\pm$0.75}}&\multirow{2}{*}{79.54\scalebox{0.4}{$\pm$0.65}}&MExMI&71.87\scalebox{0.4}{$\pm$0.88}&80.66\scalebox{0.4}{$\pm$0.82}&72.81\scalebox{0.4}{$\pm$0.70}\\
		&&&&&ActiveThief&68.86\scalebox{0.4}{$\pm$1.03}&78.42\scalebox{0.4}{$\pm$1.15}&67.70\scalebox{0.4}{$\pm$0.95}\\ \hline
	\end{tabular}
	
	\hspace{-3.3cm}\noindent Non-WM denotes the abbreviation for non-watermark.
	\vspace{-0.0cm}
\end{table}

\subsection{E.2. Removal Attack Results on Copy Models}
\label{subsec:ex_wrk_copy_models}
We evaluate the effectiveness of WRK and benchmark removal attacks on copy models extracted via MEA. Table~\ref{Table:removal_attack_copy} presents the removal performance on copy models, which complements the victim model results shown in Table~\ref{Table:removal_attack}. Compared to victim models, copy models generally exhibit similar watermark removal trends, with WRK consistently achieving low WSRs across all watermarking methods while maintaining model accuracy.

\begin{table}
	\tiny
	\renewcommand{\arraystretch}{0.97}
	\caption{Performance of WRK and benchmark removal attacks on \textbf{copy} models. Comp.\ denotes the abbreviation for composite backdoor. N/A indicates that NC does not identify any suspicious class and thus does not perform removal.}
	\label{Table:removal_attack_copy}
		\vspace{-0.235cm}
	\centering
	\begin{tabular}{>{\centering}p{0.858cm}|>{\raggedleft\arraybackslash}p{0.4cm}>{\raggedleft\arraybackslash}p{0.5cm}|>{\raggedleft\arraybackslash}p{0.4cm}>{\raggedleft\arraybackslash}p{0.5cm}|>{\raggedleft\arraybackslash}p{0.4cm}>{\raggedleft\arraybackslash}p{0.5cm}|>{\raggedleft\arraybackslash}p{0.4cm}>{\raggedleft\arraybackslash}p{0.55cm}}
		\hline
		\textbf{Removal} & \multicolumn{2}{c|}{\textbf{EWE}} & \multicolumn{2}{c|}{\textbf{MBW}} & \multicolumn{2}{c|}{\textbf{MEA-D}} & \multicolumn{2}{c}{\textbf{Blend}} \\ \hline
		\textbf{Metrics}/\% & ACC & WSR & ACC & WSR & ACC & WSR & ACC & WSR \\ \hline
		\end{tabular}
		\begin{tabular}{>{\centering}p{0.858cm}|>{\raggedleft\arraybackslash}p{0.4cm}>{\raggedleft\arraybackslash}p{0.5cm}>{\raggedleft\arraybackslash}p{0.4cm}>{\raggedleft\arraybackslash}p{0.5cm}>{\raggedleft\arraybackslash}p{0.4cm}>{\raggedleft\arraybackslash}p{0.5cm}>{\raggedleft\arraybackslash}p{0.4cm}>{\raggedleft\arraybackslash}p{0.55cm}}
		\textbf{None} 
		& 89.15\scalebox{0.5}{$\pm$0.48} & 99.65\scalebox{0.5}{$\pm$0.35} 
		& 71.32\scalebox{0.5}{$\pm$1.30} & 10.00\scalebox{0.5}{$\pm$10.0} 
		& 82.15\scalebox{0.5}{$\pm$0.31} & 99.20\scalebox{0.5}{$\pm$0.15} 
		& 89.97\scalebox{0.5}{$\pm$0.43} & 39.44\scalebox{0.5}{$\pm$12.6} \\
		\textbf{NC} 
		& 88.99\scalebox{0.5}{$\pm$0.32} & 96.67\scalebox{0.5}{$\pm$0.35} 
		& 70.33\scalebox{0.5}{$\pm$1.05} & 0.00\scalebox{0.5}{$\pm$0.00} 
		& 82.13\scalebox{0.5}{$\pm$0.35} & 75.54\scalebox{0.5}{$\pm$1.25} 
		& 88.15\scalebox{0.5}{$\pm$0.25} & \textbf{1.49}\scalebox{0.5}{$\pm$0.31} \\
		\textbf{I-BAU} 
		& 88.35\scalebox{0.5}{$\pm$0.18} & 99.15\scalebox{0.5}{$\pm$0.15} 
		& 75.16\scalebox{0.5}{$\pm$0.47} & \textbf{0.00}\scalebox{0.5}{$\pm$0.00} 
		& 83.56\scalebox{0.5}{$\pm$0.13} & 86.67\scalebox{0.5}{$\pm$1.02} 
		& 88.33\scalebox{0.5}{$\pm$0.42} & 35.85\scalebox{0.5}{$\pm$14.7} \\
		\textbf{BTI-DBF} 
		& 87.05\scalebox{0.5}{$\pm$0.23} & 10.21\scalebox{0.5}{$\pm$0.79} 
		& 70.32\scalebox{0.5}{$\pm$0.85} & \textbf{0.00}\scalebox{0.5}{$\pm$0.00} 
		& 80.21\scalebox{0.5}{$\pm$0.58} & 20.35\scalebox{0.5}{$\pm$0.65} 
		& 87.56\scalebox{0.5}{$\pm$0.28} & 3.15\scalebox{0.5}{$\pm$1.45} \\
		\textbf{CLP} 
		& 68.33\scalebox{0.5}{$\pm$12.25} & 99.98\scalebox{0.5}{$\pm$0.01} 
		& 67.67\scalebox{0.5}{$\pm$3.40} & 10.00\scalebox{0.5}{$\pm$0.00} 
		& 63.47\scalebox{0.5}{$\pm$6.35} & 98.35\scalebox{0.5}{$\pm$0.12} 
		& 71.66\scalebox{0.5}{$\pm$0.40} & 69.81\scalebox{0.5}{$\pm$5.75} \\
		\textbf{FP} 
		& 87.75\scalebox{0.5}{$\pm$0.35} & 81.67\scalebox{0.5}{$\pm$3.85} 
		& 74.35\scalebox{0.5}{$\pm$0.98} & 0.00\scalebox{0.5}{$\pm$0.00} 
		& 83.05\scalebox{0.5}{$\pm$0.14} & 62.10\scalebox{0.5}{$\pm$2.40} 
		& 88.17\scalebox{0.5}{$\pm$0.35} & 41.53\scalebox{0.5}{$\pm$9.60} \\
		\textbf{NAD} 
		& 87.46\scalebox{0.5}{$\pm$0.69} & 99.58\scalebox{0.5}{$\pm$0.22} 
		& 72.56\scalebox{0.5}{$\pm$1.02} & \textbf{0.00}\scalebox{0.5}{$\pm$0.00} 
		& 76.25\scalebox{0.5}{$\pm$0.35} & 97.40\scalebox{0.5}{$\pm$0.25} 
		& 88.17\scalebox{0.5}{$\pm$0.28} & 20.75\scalebox{0.5}{$\pm$4.55} \\
		\textbf{SEAM} 
		& 88.57\scalebox{0.5}{$\pm$0.35} & 10.23\scalebox{0.5}{$\pm$1.25} 
		& 70.59\scalebox{0.5}{$\pm$0.65} & 4.00\scalebox{0.5}{$\pm$6.00} 
		& 80.15\scalebox{0.5}{$\pm$1.13} & 35.54\scalebox{0.5}{$\pm$8.41} 
		& 88.32\scalebox{0.5}{$\pm$0.32} & 5.45\scalebox{0.5}{$\pm$3.40} \\
		\textbf{FST} 
		& 87.25\scalebox{0.5}{$\pm$0.42} & \textbf{5.31}\scalebox{0.5}{$\pm$0.24} 
		& 69.80\scalebox{0.5}{$\pm$0.34} & 4.00\scalebox{0.5}{$\pm$6.00} 
		& 79.79\scalebox{0.5}{$\pm$0.95} & 5.20\scalebox{0.5}{$\pm$1.25} 
		& 87.79\scalebox{0.5}{$\pm$0.75} & 14.15\scalebox{0.5}{$\pm$5.65} \\
		\textbf{WRK} 
		& 88.51\scalebox{0.5}{$\pm$0.39} & 5.88\scalebox{0.5}{$\pm$1.28} 
		& 71.11\scalebox{0.5}{$\pm$0.45} & 2.00\scalebox{0.5}{$\pm$8.00} 
		& 81.35\scalebox{0.5}{$\pm$0.38} & \textbf{4.76}\scalebox{0.5}{$\pm$1.22} 
		& 88.75\scalebox{0.5}{$\pm$0.27} & 2.81\scalebox{0.5}{$\pm$0.38} \\ \hline
	\end{tabular}
	\vspace{-0.42cm}
\end{table}

\subsection{E.3. Evaluation of WRK Variants}
\label{subsec:ex_wrk_variants}
We evaluate the individual impact of WRK’s two critical mechanisms: \emph{Boundary Reshaping} (BR) and the output-layer weight shift inspired by the FST framework~\cite{min2023towards}. We also compare BR against typical adversarial training (AT)~\cite{kurakin2016adversarial} to highlight its necessity. The ablation study includes two WRK variants: \emph{BR only} and \emph{FST only}, each disabling one mechanism. Table~\ref{Table:wrk_variant} presents the results.

We have the following observations. First, adversarial training (AT)~\cite{kurakin2016adversarial} is largely ineffective against most watermark and backdoor methods, with limited impact only on EWE and MBW, and consistently weaker than \emph{BR only}. Second, for EWE, MBW, and WaNet, both \emph{BR only} and full WRK achieve strong removal, indicating BR's primary role in handling noise-type triggers, while FST provides complementary benefits. In contrast, for MEA-Defender, Blend, and Composite backdoors, adding BR to FST substantially enhances removal. This indicates that BR also benefits the removal of non-noise triggers. Overall, BR and FST serve as orthogonal yet mutually beneficial components.

\begin{table}
	\tiny
	\renewcommand{\arraystretch}{1.12}
	\centering
	\caption{Performance of WRK Variants}
	\vspace{-0.25cm}
	\label{Table:wrk_variant}
	\begin{tabular}{|>{\centering}p{0.84cm} |
			>{\raggedleft\arraybackslash}p{0.38cm}>{\raggedleft\arraybackslash}p{0.445cm} |
			>{\raggedleft\arraybackslash}p{0.38cm}>{\raggedleft\arraybackslash}p{0.445cm} | 
			>{\raggedleft\arraybackslash}p{0.38cm}>{\raggedleft\arraybackslash}p{0.445cm} | 
			>{\raggedleft\arraybackslash}p{0.38cm}>{\raggedleft\arraybackslash}p{0.445cm} |}
		\hline
		\textbf{Variants}&\multicolumn{2}{>{\centering}p{1.32cm}|}{\textbf{AT}}&\multicolumn{2}{>{\centering}p{1.32cm}|}{\textbf{BR only} }&\multicolumn{2}{>{\centering}p{1.32cm}|}{\textbf{FST only}}&\multicolumn{2}{>{\centering}p{1.32cm}|}{\textbf{WRK}}\\ \hline
		\textbf{Metrics}/\%&ACC&WSR&ACC&WSR&ACC&WSR&ACC&WSR\\ \hline
		\rowcolor{gray!8}
		\multicolumn{9}{|c|}{CIFAR-10}\\ \hline
		EWE&90.49\scalebox{0.4}{$\pm$0.18}&22.89\scalebox{0.4}{$\pm$2.12}&91.32\scalebox{0.4}{$\pm$0.12}&12.30\scalebox{0.4}{$\pm$1.83}&89.96\scalebox{0.4}{$\pm$0.20}&23.55\scalebox{0.4}{$\pm$4.30}&91.44\scalebox{0.4}{$\pm$0.08}&8.75\scalebox{0.4}{$\pm$1.75}
		\\ \hline
		MBW&82.06\scalebox{0.4}{$\pm$0.25}&10.00\scalebox{0.4}{$\pm$10.0}&83.35\scalebox{0.4}{$\pm$0.22}&0.00\scalebox{0.4}{$\pm$0.00}&83.17\scalebox{0.4}{$\pm$0.12}&6.00\scalebox{0.4}{$\pm$6.00}&82.50\scalebox{0.4}{$\pm$0.15}&4.00\scalebox{0.4}{$\pm$6.00}
		\\ \hline
		MEA-D&85.90\scalebox{0.4}{$\pm$0.14}&81.44\scalebox{0.4}{$\pm$3.21}&86.50\scalebox{0.4}{$\pm$0.17}&72.15\scalebox{0.4}{$\pm$2.82}&86.08\scalebox{0.4}{$\pm$0.08}&6.78\scalebox{0.4}{$\pm$0.30}&85.46\scalebox{0.4}{$\pm$0.12}&7.64\scalebox{0.4}{$\pm$0.28} 
		\\\hline
		Blend &92.25\scalebox{0.4}{$\pm$0.12}&85.02\scalebox{0.4}{$\pm$4.12}&92.11\scalebox{0.4}{$\pm$0.14}&45.20\scalebox{0.4}{$\pm$3.65}&92.48\scalebox{0.4}{$\pm$0.08}&57.51\scalebox{0.4}{$\pm$6.75}&92.40\scalebox{0.4}{$\pm$0.07}&9.80\scalebox{0.4}{$\pm$2.35} 
		\\ \hline
		WaNet&91.45\scalebox{0.4}{$\pm$0.12}&83.54\scalebox{0.4}{$\pm$3.85}&92.31\scalebox{0.4}{$\pm$0.11}&5.65\scalebox{0.4}{$\pm$0.90}&91.88\scalebox{0.4}{$\pm$0.08}&81.01\scalebox{0.4}{$\pm$5.40}&92.19\scalebox{0.4}{$\pm$0.05}&0.56\scalebox{0.4}{$\pm$0.25}
		\\ \hline
		Composite&91.21\scalebox{0.4}{$\pm$0.16}&90.35\scalebox{0.4}{$\pm$4.45}&91.19\scalebox{0.4}{$\pm$0.19}&77.20\scalebox{0.4}{$\pm$3.22}&91.07\scalebox{0.4}{$\pm$0.20}&3.80\scalebox{0.4}{$\pm$0.50}&91.82\scalebox{0.4}{$\pm$0.15}&6.00\scalebox{0.4}{$\pm$0.30}
		\\ \hline
		\rowcolor{gray!8}
		\multicolumn{9}{|c|}{ImageNette}\\ \hline
		EWE&86.15\scalebox{0.4}{$\pm$0.73}&84.89\scalebox{0.4}{$\pm$3.82}&86.01\scalebox{0.4}{$\pm$0.64}&19.21\scalebox{0.4}{$\pm$2.12}&84.03\scalebox{0.4}{$\pm$0.63}&69.14\scalebox{0.4}{$\pm$10.3}&84.35\scalebox{0.4}{$\pm$0.37}&10.55\scalebox{0.4}{$\pm$0.74}
		\\ \hline
		MBW&79.14\scalebox{0.4}{$\pm$0.89}&6.00\scalebox{0.4}{$\pm$6.00}&77.94\scalebox{0.4}{$\pm$1.04}&4.00\scalebox{0.4}{$\pm$6.00} &78.85\scalebox{0.4}{$\pm$1.10}&8.00\scalebox{0.4}{$\pm$8.00}&79.39\scalebox{0.4}{$\pm$0.95}&6.00\scalebox{0.4}{$\pm$6.00}
		\\ \hline
		MEA-D&74.15\scalebox{0.4}{$\pm$0.81}&64.35\scalebox{0.4}{$\pm$9.12}&73.53\scalebox{0.4}{$\pm$0.74}&51.03\scalebox{0.4}{$\pm$2.92}&72.93\scalebox{0.4}{$\pm$0.86}&36.84\scalebox{0.4}{$\pm$6.46}&74.06\scalebox{0.4}{$\pm$0.53}&3.62\scalebox{0.4}{$\pm$0.81}
		\\\hline
		Blend &81.14\scalebox{0.4}{$\pm$0.52}&99.23\scalebox{0.4}{$\pm$0.30}&80.45\scalebox{0.4}{$\pm$0.56}&87.90\scalebox{0.4}{$\pm$3.12}&80.35\scalebox{0.4}{$\pm$0.92}&6.35\scalebox{0.4}{$\pm$1.85}&80.56\scalebox{0.4}{$\pm$0.45}&3.53\scalebox{0.4}{$\pm$0.91}
		\\ \hline
		WaNet&87.05\scalebox{0.4}{$\pm$0.64}&81.43\scalebox{0.4}{$\pm$3.71}&86.35\scalebox{0.4}{$\pm$0.72}&12.51\scalebox{0.4}{$\pm$1.48}&85.24\scalebox{0.4}{$\pm$0.75}&0.83\scalebox{0.4}{$\pm$0.83}&86.89\scalebox{0.4}{$\pm$0.30}&0.25\scalebox{0.4}{$\pm$0.25}
		\\ \hline
		Composite&85.81\scalebox{0.4}{$\pm$0.78}&92.84\scalebox{0.4}{$\pm$3.95}&85.42\scalebox{0.4}{$\pm$0.76}&74.99\scalebox{0.4}{$\pm$3.02}&84.61\scalebox{0.4}{$\pm$0.80}&0.59\scalebox{0.4}{$\pm$0.35}&86.19\scalebox{0.4}{$\pm$0.25}&5.96\scalebox{0.4}{$\pm$0.85}
		\\ \hline
	\end{tabular}
\end{table}

\subsection{E.4. Sensitivity Analysis of WRK Hyperparameters}
\label{subsec:wrk_hyperparameter}
We evaluate the sensitivity of WRK to its two key hyperparameters. The first is $\epsilon$, which controls the magnitude of adversarial perturbations used in boundary reshaping. The second is $\alpha$, which is the coefficient of the parameter-shift regularization term in Eq.~\eqref{eq:wrk_loss}. We vary $\epsilon \in \{0.01, 0.03, 0.05\}$ and $\alpha \in \{0.05, 1.0, 1.5\}$ to assess their impact on model accuracy (ACC) and watermark success rate (WSR) across EWE and MEA-Defender watermarks.

Table~\ref{Table:wrk_hyperparameter} presents the results. For EWE, $\epsilon = 0.03$ with $\alpha = 1.0$ achieves optimal balance. This configuration reduces WSR to $8.75\%$ and maintains $91.44\%$ accuracy. For MEA-Defender, larger $\epsilon$ is required. At $\epsilon = 0.03$, $\alpha = 1.5$ yields WSR of $7.64\%$.  Overall, WRK demonstrates moderate sensitivity within the tested ranges.

\begin{table}
	\tiny
	\renewcommand{\arraystretch}{1.12}
	\centering
	\caption{Sensitivity Analysis of WRK Hyperparameters}
	\vspace{-0.25cm}
	\label{Table:wrk_hyperparameter}
	\begin{tabular}{|>{\centering}p{0.75cm} |
			>{\centering}p{0.35cm} |
			>{\raggedleft\arraybackslash}p{0.5cm}>{\raggedleft\arraybackslash}p{0.7cm} | 
			>{\raggedleft\arraybackslash}p{0.5cm}>{\raggedleft\arraybackslash}p{0.7cm} | 
			>{\raggedleft\arraybackslash}p{0.5cm}>{\raggedleft\arraybackslash}p{0.7cm} |}
		\hline
		\textbf{WM}&\textbf{$\epsilon$}&\multicolumn{2}{>{\centering}p{1.1cm}|}{$\alpha = 0.05$}&\multicolumn{2}{>{\centering}p{1.1cm}|}{$\alpha = 1.0$}&\multicolumn{2}{>{\centering}p{1.1cm}|}{$\alpha = 1.5$}\\ \hline
		\multicolumn{2}{|c|}{\textbf{Metric/\%}}&ACC&WSR&ACC&WSR&ACC&WSR\\ \hline
		\multirow{3}{*}{EWE}&0.01&91.49\scalebox{0.4}{$\pm$0.15}&48.45\scalebox{0.4}{$\pm$7.64}&91.12\scalebox{0.4}{$\pm$0.18}&30.95\scalebox{0.4}{$\pm$9.83}&91.82\scalebox{0.4}{$\pm$0.08}&90.99\scalebox{0.4}{$\pm$0.15}\\
		&0.03&91.04\scalebox{0.4}{$\pm$0.10}&6.50\scalebox{0.4}{$\pm$2.39}&91.44\scalebox{0.4}{$\pm$0.08}&8.75\scalebox{0.4}{$\pm$1.75}&91.16\scalebox{0.4}{$\pm$0.14}&8.31\scalebox{0.4}{$\pm$8.71}\\
		&0.05&89.14\scalebox{0.4}{$\pm$0.41}&2.03\scalebox{0.4}{$\pm$0.25}&88.19\scalebox{0.4}{$\pm$1.02}&0.86\scalebox{0.4}{$\pm$0.08}&89.51\scalebox{0.4}{$\pm$0.79}&4.02\scalebox{0.4}{$\pm$1.36}\\
		\hline
		\multirow{3}{*}{MEA-D}&0.01&86.60\scalebox{0.4}{$\pm$0.09}&81.20\scalebox{0.4}{$\pm$2.32}&86.91\scalebox{0.4}{$\pm$0.12}&82.38\scalebox{0.4}{$\pm$3.23}&86.29\scalebox{0.4}{$\pm$0.04}&74.64\scalebox{0.4}{$\pm$4.84}\\
		&0.03&86.44\scalebox{0.4}{$\pm$0.11}&50.92\scalebox{0.4}{$\pm$2.34}&85.02\scalebox{0.4}{$\pm$2.02}&15.20\scalebox{0.4}{$\pm$9.86}&85.46\scalebox{0.4}{$\pm$0.12}&7.64\scalebox{0.4}{$\pm$0.28}\\
		&0.05&79.93\scalebox{0.4}{$\pm$6.34}&24.57\scalebox{0.4}{$\pm$10.21}&80.08\scalebox{0.4}{$\pm$10.76}&16.01\scalebox{0.4}{$\pm$16.00}&80.57\scalebox{0.4}{$\pm$6.48}&8.91\scalebox{0.4}{$\pm$3.63}\\
		\hline
	\end{tabular}
\end{table}

\subsection{E.5. Watermark Decoupling Curves on Victim Models}
\begin{figure}
	\centering{
		\hspace{-0.23cm}\subfloat[Victim Model]{
			\label{subfig:cfw_a}
			\includegraphics[width=0.319\columnwidth]{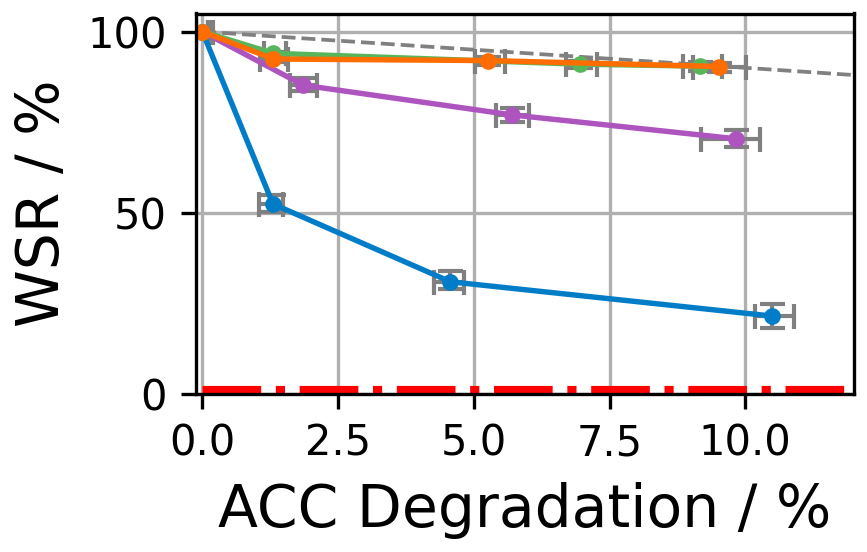}
		}	\hspace{-0.15cm}
		\subfloat[Victim Model]{
			\label{subfig:cfw_b}
			\includegraphics[width=0.319\columnwidth]{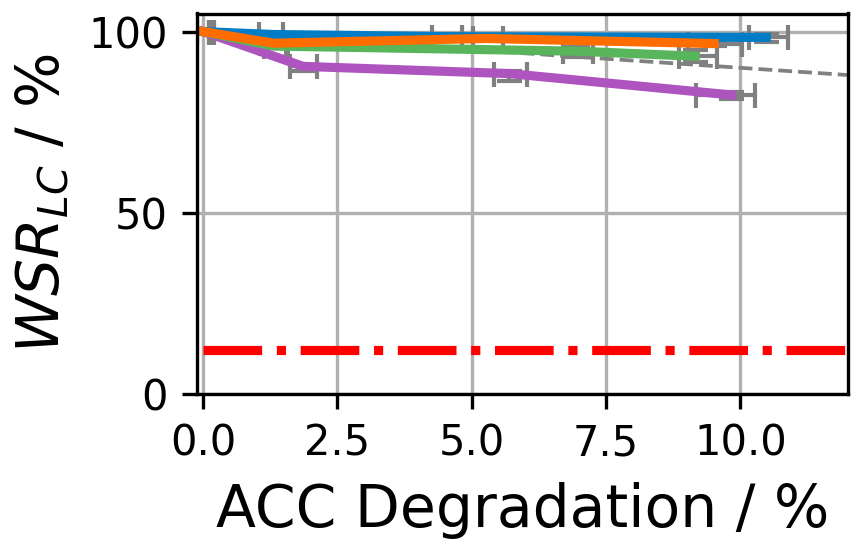}
		}\hspace{-0.15cm}
		\subfloat[Victim Model]{
			\label{subfig:cfw_c}
			\includegraphics[width=0.319\columnwidth]{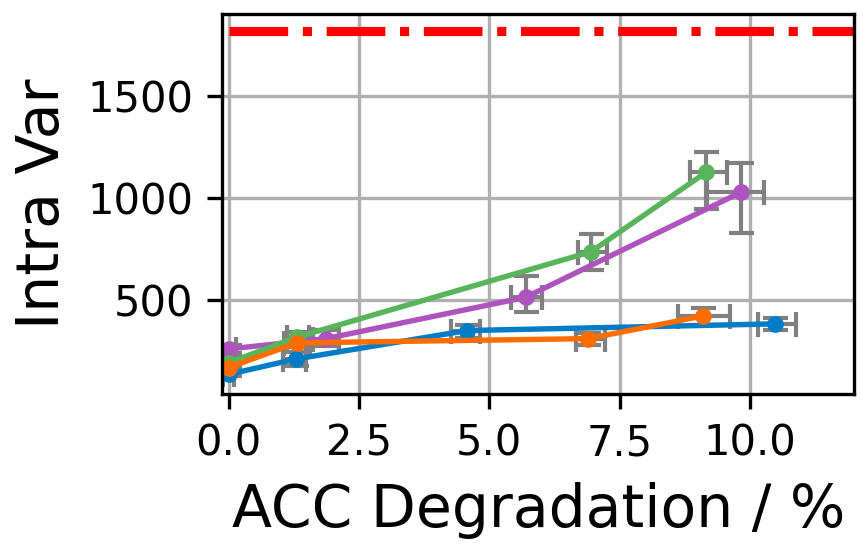}
		}\hspace{-0.1cm}\\
		\includegraphics[width=1.0\columnwidth]{cfw_decoupling_legend.png}
		\vspace{-0.25cm}
		\caption{Watermark decoupling curves of CFW on victim models. Vertical lines indicate error bars.}
		\label{app:fig_cfw_decoupling}
	}
\end{figure}

Figure~\ref{app:fig_cfw_decoupling} presents supplementary results showing the decoupling behavior of Class-Feature Watermarks (CFW) on victim models under WRK removal. 

\subsection{E.6. Comparing WSR and $\text{WSR}_{\text{LC}}$ for Watermark Baselines and CFW}
\label{subsec:wsr_comparison}
We compare the resilience of WSR and $\text{WSR}_{\text{LC}}$ between CFW and representative baseline watermarks. Recall that $\text{WSR}_{\text{LC}}$ is defined as the sum of WSRs over watermark and deformation labels, providing clustering-based evidence for ownership verification.

Table~\ref{Table:cfw_wsr} reports WSR values for CFW across five datasets. On victim models, CFW maintains perfect WSR before WRK removal. After WRK attacks, WSR remains above $92\%$ on victim models. On copy models extracted via MEA, WSR ranges from $82.67\%$ to $93.12\%$ before removal, and drops to between $56.00\%$ and $73.45\%$ after WRK removal. 

Table~\ref{Table:baseline_wsr_lc} reports $\text{WSR}_{\text{LC}}$ values for representative baseline watermarks. On victim models, all baselines achieve perfect $\text{WSR}_{\text{LC}}$ before removal. After WRK attacks, however, $\text{WSR}_{\text{LC}}$ drops dramatically to between $15.52\%$ and $22.83\%$ on victim models and between $15.52\%$ and $22.34\%$ on copy models. This indicates that baseline watermarks fail to maintain clustering structure under removal attacks.

When compared CFW and watermark baselines, CFW demonstrates superior resilience in terms of $\text{WSR}_{\text{LC}}$ and WSR. While baseline watermarks suffer severe degradation in $\text{WSR}_{\text{LC}}$ and WSR after WRK removal, CFW maintains high values that reflect its robust clustering structure.

\begin{table}
	\tiny
	\renewcommand{\arraystretch}{1.12}
	\centering
	\caption{WSR Performance of CFW Across Datasets}
	\vspace{-0.25cm}
	\label{Table:cfw_wsr}
	\begin{tabular}{|>{\centering}p{2.2cm} |
			>{\raggedleft\arraybackslash}p{0.99cm} | 
			>{\raggedleft\arraybackslash}p{0.99cm} | 
			>{\raggedleft\arraybackslash}p{0.99cm} | 
			>{\raggedleft\arraybackslash}p{0.99cm} |}
		\hline
		\textbf{Dataset}&\multicolumn{2}{c|}{\textbf{Victim Model}}&\multicolumn{2}{c|}{\textbf{Copy Model}}\\ \hline
		\textbf{Removal}&None&WRK&None&WRK\\ \hline
		CIFAR-10&100.00\scalebox{0.4}{$\pm$0.00}&94.72\scalebox{0.4}{$\pm$0.67}&91.20\scalebox{0.4}{$\pm$1.33}&65.60\scalebox{0.4}{$\pm$2.00}\\
		\hline
		CIFAR-20&100.00\scalebox{0.4}{$\pm$0.00}&94.00\scalebox{0.4}{$\pm$0.67}&82.67\scalebox{0.4}{$\pm$1.30}&56.00\scalebox{0.4}{$\pm$1.33}\\
		\hline
		Imagenette&100.00\scalebox{0.4}{$\pm$0.00}&93.23\scalebox{0.4}{$\pm$2.36}&82.67\scalebox{0.4}{$\pm$2.20}&61.33\scalebox{0.4}{$\pm$2.57}\\
		\hline
		DBPedia&100.00\scalebox{0.4}{$\pm$0.00}&92.15\scalebox{0.4}{$\pm$0.45}&93.12\scalebox{0.4}{$\pm$0.96}&73.45\scalebox{0.4}{$\pm$1.33}\\
		\hline
		Speech Commands&98.32\scalebox{0.4}{$\pm$0.15}&93.33\scalebox{0.4}{$\pm$0.67}&92.35\scalebox{0.4}{$\pm$0.64}&70.12\scalebox{0.4}{$\pm$0.64}\\
		\hline
	\end{tabular}
\end{table}

\begin{table}
	\tiny
	\renewcommand{\arraystretch}{1.12}
	\centering
	\caption{$\text{WSR}_{\text{LC}}$ Performance of Representative Baseline Watermarks}
	\vspace{-0.25cm}
	\label{Table:baseline_wsr_lc}
	\begin{tabular}{|>{\centering}p{2.2cm} |
			>{\raggedleft\arraybackslash}p{0.99cm} | 
			>{\raggedleft\arraybackslash}p{0.99cm} | 
			>{\raggedleft\arraybackslash}p{0.99cm} | 
			>{\raggedleft\arraybackslash}p{0.99cm} |}
		\hline
		\textbf{WM}&\multicolumn{2}{c|}{\textbf{Victim Model}}&\multicolumn{2}{c|}{\textbf{Copy Model}}\\ \hline
		\textbf{Removal}&None&WRK&None&WRK\\ \hline
		EWE&100.00\scalebox{0.4}{$\pm$0.00}&19.89\scalebox{0.4}{$\pm$1.47}&99.65\scalebox{0.4}{$\pm$0.35}&22.34\scalebox{0.4}{$\pm$0.63}\\
		\hline
		MEA-D&100.00\scalebox{0.4}{$\pm$0.00}&19.90\scalebox{0.4}{$\pm$0.35}&99.53\scalebox{0.4}{$\pm$0.04}&15.52\scalebox{0.4}{$\pm$1.03}\\
		\hline
		Blend&98.43\scalebox{0.4}{$\pm$0.35}&22.83\scalebox{0.4}{$\pm$4.32}&50.86\scalebox{0.4}{$\pm$0.72}&16.49\scalebox{0.4}{$\pm$0.79}\\
		\hline
	\end{tabular}
\end{table}

\subsection{E.7. Resilience of CFW to Additional Removal Attacks}
\label{app:sub_additional_attack}
Table~\ref{Table:removal_attack_cfw} compares CFW with nine popular removal techniques and WRK removal on CIFAR‑10 victim models and their MExMI copies. CLP occasionally causes accuracy degradation exceeding $2\%$, which occurs when at least one neuron must be pruned. From Table~\ref{Table:removal_attack_cfw}, we have the following key observations. First, CFW exhibits strong resilience across all evaluated removal attacks, maintaining $\text{WSR}_{\text{LC}} \geq 96\%$ on victim models and over $72\%$ on copy models. Second, the intra-class variance remains extremely low even after the attack, providing a strong clustering signal for ownership verification. Specifically, variance values are consistently less than one-third of those in non-watermarked models. These results highlight that CFW resists diverse removal strategies by combining class-level watermarking with clustering-based verification.

\begin{table}
	\tiny
	\renewcommand{\arraystretch}{1.12}
	\caption{Resilience of CFW against Other Removal Attacks}
	\vspace{-0.25cm}
	\label{Table:removal_attack_cfw}
	\centering
	\begin{tabular}{| >{\centering}p{1.405cm} | >{\raggedleft\arraybackslash}p{0.5cm}>{\raggedleft\arraybackslash}p{0.52cm}  >{\raggedleft\arraybackslash}p{0.96cm}|>{\raggedleft\arraybackslash}p{0.5cm}  >{\raggedleft\arraybackslash}p{0.52cm}>{\raggedleft\arraybackslash}p{0.96cm} |}
		\hline
		\rowcolor{gray!8}
		\textbf{Removal}&\multicolumn{3}{c|}{\textbf{Victim Model}}&\multicolumn{3}{c|}{\textbf{Copy Model} (MExMI)}\\ \hline
		\rowcolor{gray!8}
		\textbf{Metrics}/\%&ACC&$\text{WSR}_{\text{LC}}$&Var(\tiny$\times10^2$)\tiny$\downarrow$&ACC&$\text{WSR}_{\text{LC}}$&Var(\tiny$\times10^2$)\tiny$\downarrow$\\ \hline
		Non-Watermark&93.55\scalebox{0.50}{$\pm$0.19}&12.00\scalebox{0.50}{$\pm$2.67}&18.18\scalebox{0.50}{$\pm$4.25}&89.81\scalebox{0.50}{$\pm$0.95}&11.33\scalebox{0.50}{$\pm$2.67}&19.78\scalebox{0.50}{$\pm$4.75}\\ \hline
		Non-Removal&93.26\scalebox{0.50}{$\pm$0.12}&100.00\scalebox{0.50}{$\pm$0.00}&1.68\scalebox{0.50}{$\pm$0.45}&89.29\scalebox{0.50}{$\pm$1.06}&94.00\scalebox{0.50}{$\pm$1.33}&2.74\scalebox{0.50}{$\pm$0.95}\\ \hline
		NC&92.38\scalebox{0.50}{$\pm$0.25}&98.67\scalebox{0.50}{$\pm$1.33}&1.66\scalebox{0.50}{$\pm$0.42}&89.04\scalebox{0.50}{$\pm$1.12}&72.67\scalebox{0.50}{$\pm$2.67}&5.15\scalebox{0.50}{$\pm$1.85}\\ \hline
		I-BAU&91.49\scalebox{0.50}{$\pm$0.30}&98.67\scalebox{0.50}{$\pm$1.33}&1.77\scalebox{0.50}{$\pm$0.50}&89.04\scalebox{0.50}{$\pm$0.98}&80.00\scalebox{0.50}{$\pm$2.00}&3.20\scalebox{0.50}{$\pm$1.25}\\ \hline
		BTI-DBF&91.57\scalebox{0.50}{$\pm$0.28}&96.67\scalebox{0.50}{$\pm$2.67}&1.66\scalebox{0.50}{$\pm$0.48}&88.15\scalebox{0.50}{$\pm$1.51}&72.67\scalebox{0.50}{$\pm$2.67}&5.23\scalebox{0.50}{$\pm$1.95}\\ \hline
		CLP&87.44\scalebox{0.50}{$\pm$0.75}&99.33\scalebox{0.50}{$\pm$0.67}&2.00\scalebox{0.50}{$\pm$0.65}&87.01\scalebox{0.50}{$\pm$0.98}&90.00\scalebox{0.50}{$\pm$2.00}&4.20\scalebox{0.50}{$\pm$1.65}\\ \hline
		Fine pruning&84.20\scalebox{0.50}{$\pm$1.25}&100.00\scalebox{0.50}{$\pm$0.00}&1.78\scalebox{0.50}{$\pm$0.60}&67.33\scalebox{0.50}{$\pm$1.02}&96.67\scalebox{0.50}{$\pm$1.33}&3.16\scalebox{0.50}{$\pm$1.35}\\ \hline
		NAD&91.76\scalebox{0.50}{$\pm$0.22}&98.00\scalebox{0.50}{$\pm$2.00}&1.81\scalebox{0.50}{$\pm$0.55}&88.20\scalebox{0.50}{$\pm$0.94}&73.33\scalebox{0.50}{$\pm$2.67}&4.87\scalebox{0.50}{$\pm$1.75}\\ \hline
		AT&92.15\scalebox{0.50}{$\pm$0.20}&99.33\scalebox{0.50}{$\pm$0.67}&1.64\scalebox{0.50}{$\pm$0.40}&89.23\scalebox{0.50}{$\pm$1.23}&73.33\scalebox{0.50}{$\pm$2.67}&4.66\scalebox{0.50}{$\pm$1.65}\\ \hline
		SEAM&92.35\scalebox{0.50}{$\pm$0.18}&100.00\scalebox{0.50}{$\pm$0.00}&1.72\scalebox{0.50}{$\pm$0.52}&88.84\scalebox{0.50}{$\pm$1.41}&80.00\scalebox{0.50}{$\pm$2.00}&3.54\scalebox{0.50}{$\pm$1.45}\\ \hline
		FST&91.21\scalebox{0.50}{$\pm$0.32}&96.00\scalebox{0.50}{$\pm$2.00}&1.61\scalebox{0.50}{$\pm$0.45}&89.02\scalebox{0.50}{$\pm$0.98}&78.67\scalebox{0.50}{$\pm$2.67}&2.85\scalebox{0.50}{$\pm$1.15}\\ \hline
		WRK&91.95\scalebox{0.50}{$\pm$0.23}&96.67\scalebox{0.50}{$\pm$2.67}&2.89\scalebox{0.50}{$\pm$0.95}&88.37\scalebox{0.50}{$\pm$1.13}&79.33\scalebox{0.50}{$\pm$2.67}&4.90\scalebox{0.50}{$\pm$1.85}\\ \hline
	\end{tabular}
	
	\hspace{-2.65cm} Arrows represent the trend toward better watermark performance.
	\vspace{-0.0cm}
\end{table}

\subsection{E.8. Impact of $\text{CD}^2$ on CFW Stability}
\begin{figure}
	\vspace{-0.05cm}
	\centering{
		\subfloat[CFW]{
			\label{subfig:ex_cf_dppc}
			\hspace{-0.3cm}\includegraphics[width=0.415\columnwidth]{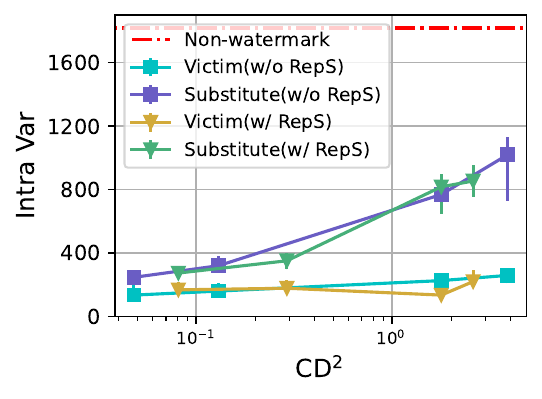}
		}\qquad
		\subfloat[WRK-attacked CFW]{
			\includegraphics[width=0.415\columnwidth]{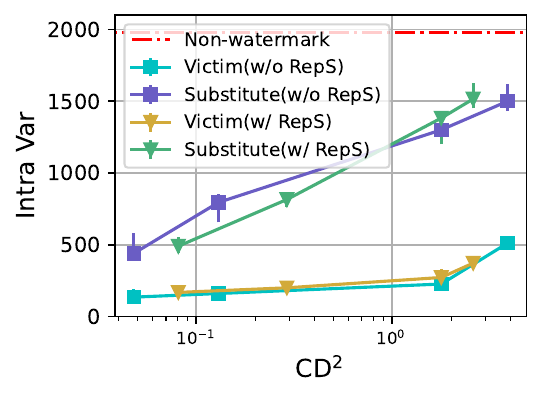}
			\label{subfig:ex_cf_dppc_wrk}
		}
		\vspace{-0.15cm}
		\caption{$\text{CD}^2$ versus Intra-class Variance. The vertical lines represent error bars.}
		\label{fig:ex_dppc}}
		\vspace{-0.3cm}
\end{figure}

This section evaluates the relationship between $\text{CD}^2$ and the clustering stability of CFW, which is measured by intra-class variance. Here, two scenarios are observed: one without an attack and the other with WRK attacks. Each experiment is further divided into two setup conditions: one with RepS and one without. The experiments are conducted on CIFAR-10. Figure~\ref{fig:ex_dppc} visualizes these relationships for each scenario.

First, we observe that $\text{CD}^2$ optimization significantly affects the clustering stability of the copy model. When $\text{CD}^2$ loss (calculated with Equation~\ref{eq:loss_cd2_final_main}) is below $0.1$, both the MEA-post and MEA-removal-post stability of the copy model achieve favorable conditions, with consistently low variance. However, as $\text{CD}^2$ increases, the MEA-post stability (Figure~\ref{subfig:ex_cf_dppc}) and removal-post stability (Figure~\ref{subfig:ex_cf_dppc_wrk}) decline rapidly. These findings indicate that, in terms of representation space clustering, CFW becomes unstable to MEA and removal attacks if $\text{CD}^2$ is not properly optimized. Finally, RepS does not have a clear impact on the stability of watermark tasks.

\subsection{E.9. Impact of Representation Entanglement (RE) of CFW on MEA Transferability}
\begin{figure}
	\centering{
		\subfloat[w/o $\text{CD}^2$]{
			\label{subfig:ex_cf_reps_wo_pd3}
			\includegraphics[width=0.39\columnwidth]{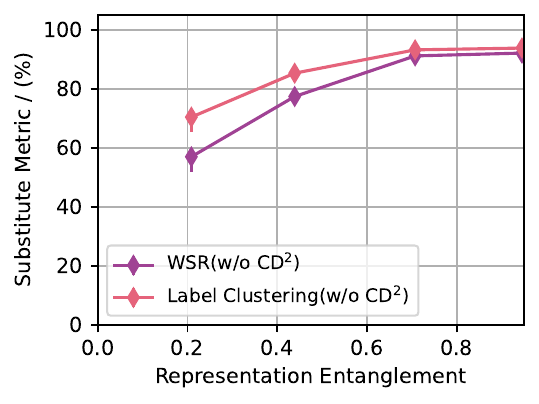}
		}\qquad
		\subfloat[w/ $\text{CD}^2$]{
			\label{subfig:ex_cf_reps_w_pd3}
			\includegraphics[width=0.39\columnwidth]{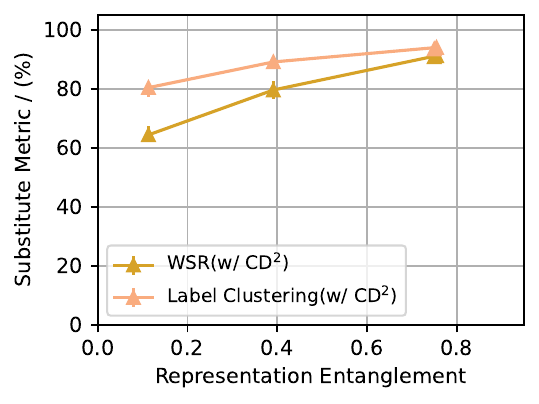}
		}
	}
	\vspace{-0.15cm}
	\caption{MEA transferability versus the representation entanglement ($\mathcal{RE}$). The vertical lines represent error bars.}
	\vspace{-0.3cm}
	\label{fig:ex_maxmum_orth_vs_mea_trans}
\end{figure}

This section investigates the relationship between the representation entanglement ($\mathcal{RE}$) and MEA transferability for CFW, evaluated through the copy model's WSR and $\text{WSR}_\text{LC}$. Two sets of experiments are conducted: one with $\text{CD}^2$ optimization and one without, where $\mathcal{RE}$ is controlled by the RepS coefficient. The experiments use the CIFAR-10 dataset. The results presented in Figure~\ref{fig:ex_maxmum_orth_vs_mea_trans} reveal several key findings. First, when $\mathcal{RE}$ is larger than $0.3$, MEA transferability consistently reaches optimal levels. Additionally, even when $\mathcal{RE}$ exceeds $0.8$, WSR remains above $50\%$, primarily because CFW is constructed with real-life samples. Lastly, while $\text{CD}^2$ has limited impact on WSR, it significantly enhances $\text{WSR}_\text{LC}$, achieving over $80\%$ even when $\mathcal{RE}<0.2$. This improvement is attributed to $\text{CD} ^2$'s ability to preserve clustering stability during MEA.

\subsection{E.10. Ablation Study: The Impact of Copy Model's Architectures on Class-feature Watermarks}
\label{subsec:ex_arch}
\begin{figure}
	\centering{
		\subfloat[Label Clustering]{
			\includegraphics[width=0.3\columnwidth]{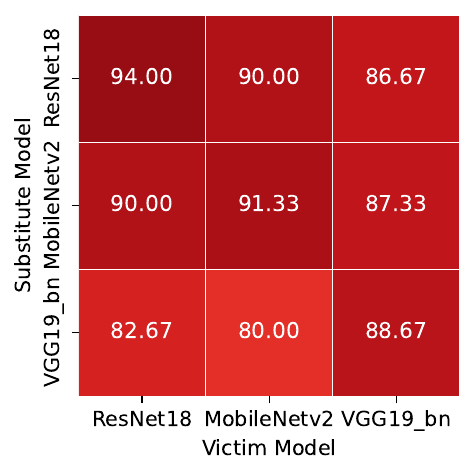} }
		\subfloat[WRK-Attacked Label Clustering]{
			\includegraphics[width=0.3\columnwidth]{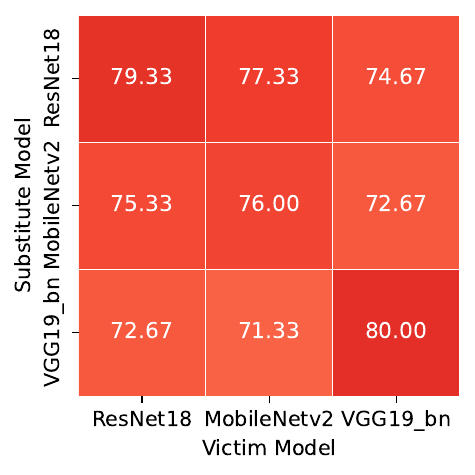} }
		\subfloat[MEA Accuracy]{
			\includegraphics[width=0.345\columnwidth]{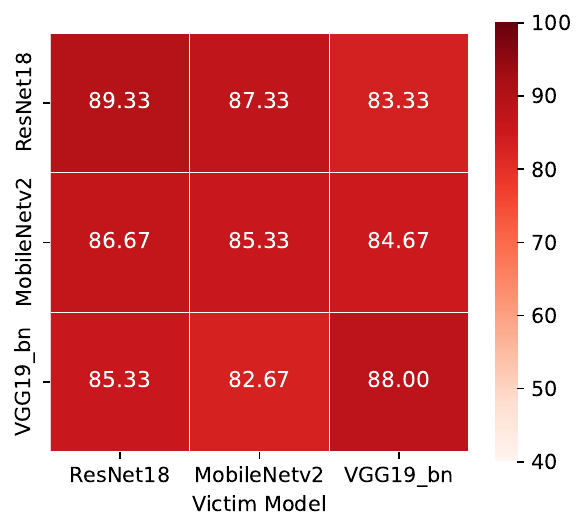} }
		\vspace{-0.1cm}
		\caption{Performance of CFW with different architectures used in model extraction attacks. The horizontal labels are victim models, and the vertical labels are copy models.}
		\label{fig:arch_heat}
	}
\end{figure}

In model extraction attacks, the copy model's architecture used by the adversary usually differs from that of the victim model. Therefore, we study the consistency of CFW across different architectures. Specifically, we cross-test three architectures: ResNet18, MobileNet, and VGG19-bn on CIFAR-10. In this setup, the victim and copy models are assigned different architectures in each combination, and Figure~\ref{fig:arch_heat} presents CFW's performance using heatmap grids. The results indicate that CFW's performance is minimally affected by model architecture, displaying high consistency. This is because the CFW functions as a task and is independent of the underlying architectures.

\subsection{E.11. Discussions}
\label{sec:discussion}
\noindent\textbf{Piracy Attack}. A piracy attack occurs when the adversaries embed their watermarks in a stolen model to evade watermark detection~\cite{liu2024false}. This creates ambiguity in ownership verification, as two watermarks exist simultaneously. To prevent such ownership confusion, the simplest countermeasure is to timestamp the watermark model, samples, and verifier through a trusted and accessible platform, such as privately uploading them to an open-source repository. This ensures that even if the adversary performs a piracy attack, they cannot predate the defender's verified timestamp.

We experimentally evaluate CFW's resilience against such piracy attacks under two adaptive injection scenarios. In the same-label scenario, the adversary injects watermarks targeting the same label as CFW. In the cross-label scenario, the adversary targets a different label. Table~\ref{Table:false_claim} presents the experimental results on CIFAR-10. The results indicates that even with injected watermarks, CFW maintains strong verification capability. On victim models, CFW achieves perfect $\text{WSR}_{\text{LC}}$ of $100\%$ under both same-label and cross-label injection attacks. On copy models, CFW maintains $\text{WSR}_{\text{LC}}$ above $92.70\%$ in both scenarios. In contrast, the injected piracy watermarks achieve significantly lower $\text{WSR}_{\text{LC}}$ values, ranging from $10.00\%$ to $62.50\%$. This demonstrates that CFW's clustering-based verification remains robust even when competing watermarks are present. 

\begin{table}
	\tiny
	\renewcommand{\arraystretch}{1.12}
	\centering
	\caption{CFW Resilience Against False Claim Attacks via Adaptive Watermark Injection}
	\vspace{-0.25cm}
	\label{Table:false_claim}
	\begin{tabular}{|>{\centering}p{1.2cm} |
			>{\centering}p{0.8cm} |
			>{\raggedleft\arraybackslash}p{0.85cm} | 
			>{\raggedleft\arraybackslash}p{1.65cm} | 
			>{\raggedleft\arraybackslash}p{1.55cm} |}
		\hline
		\textbf{Target Label}&\textbf{Model}&\textbf{ACC}/\%&\textbf{Piracy $\text{WSR}_{\text{LC}}$}/\%&\textbf{CFW $\text{WSR}_{\text{LC}}$}/\%\\ \hline
		\multirow{2}{*}{Same}&Victim&90.79\scalebox{0.4}{$\pm$0.68}&62.50\scalebox{0.4}{$\pm$8.18}&100.00\scalebox{0.4}{$\pm$0.00}\\ \cline{2-5}
		&Copy&87.14\scalebox{0.4}{$\pm$0.87}&57.60\scalebox{0.4}{$\pm$9.31}&97.70\scalebox{0.4}{$\pm$0.17}\\
		\hline
		\multirow{2}{*}{Cross}&Victim&90.69\scalebox{0.4}{$\pm$2.53}&10.00\scalebox{0.4}{$\pm$3.58}&100.00\scalebox{0.4}{$\pm$0.00}\\ \cline{2-5}
		&Copy&86.85\scalebox{0.4}{$\pm$4.43}&32.00\scalebox{0.4}{$\pm$11.85}&92.70\scalebox{0.4}{$\pm$2.86}\\
		\hline
	\end{tabular}
\end{table}

\begin{figure}
	\centering{
		\subfloat[Victim Model]{
			\includegraphics[width=0.425\columnwidth]{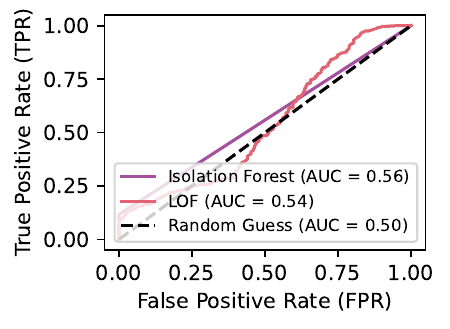} }\quad
		\subfloat[Copy Model (MExMI)]{
			\includegraphics[width=0.425\columnwidth]{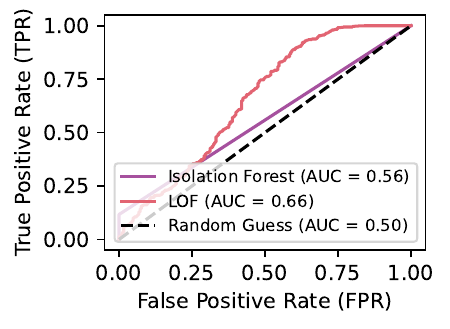} }
	}
	\vspace{-0.1cm}
	\caption{AUC results of abnormal detections on CFW}
	\label{fig:ab_detect}
	\vspace{-0.0cm}
\end{figure}

\noindent\textbf{Watermark Detection.}
The section titled \emph{Experiments for Class-feature Watermarks} has yet to discuss the stealthiness property (\textbf{Prop.6}). When the stolen model is deployed online, adversaries may filter out potential watermark data and refuse to provide correct query labels to evade watermark verification. To assess CFW's stealthiness, we evaluate two anomaly detection methods: Local Outlier Factor (LOF)~\cite{breunig2000lof} and Isolation Forest~\cite{liu2008isolation}, applied to the last hidden layer, following EWE~\cite{jia2021entangled}. Since CFW relies on clustering in the representation space, these methods infer watermark queries as high-density points, contrary to their original detection principles. Figure~\ref{fig:ab_detect} shows AUC results on the CIFAR-10 dataset with 1,200 queries. The highest AUC reached only $0.66$, indicating that watermark queries are difficult to distinguish.

